\documentclass[11pt]{article}
\usepackage[T1]{fontenc}
\usepackage{babel}
\usepackage{verbatim}
\usepackage{bm}
\usepackage{amsmath,amstext,amsthm,amssymb}
\usepackage{thmtools}
\usepackage{thm-restate}
\usepackage{standalone}
\usepackage{float}
\usepackage{bm}
\usepackage{bbm}
\usepackage{xspace}
\usepackage{graphicx}
\usepackage{mathtools}
\usepackage{subcaption,array}
\usepackage{wrapfig}
\usepackage{booktabs}
\usepackage{makecell}
\usepackage{array}
\usepackage{multirow}
\usepackage[usenames,dvipsnames]{xcolor}
\usepackage{tikz}
\usetikzlibrary{arrows.meta}
\usepackage{wrapfig}
\usepackage{boxedminipage,wrapfig}
\usepackage{pifont}%

\usepackage[letterpaper,top=1in,bottom=2cm,left=1in,right=1in,marginparwidth=1.75cm,footskip=1cm]{geometry}
\usepackage[font={small,it}]{caption}
\usepackage{amssymb}
\usepackage{amsthm}
\usepackage{comment}
\usepackage{mdframed}
\usepackage{xspace}
\usepackage{multicol}
\usepackage{tabularx}
\usepackage{makecell}
\usepackage{parskip}

\usepackage[ruled,vlined,linesnumbered,algonl,procnumbered]{algorithm2e}
\SetEndCharOfAlgoLine{}
\SetKwComment{Comment}{\footnotesize$\triangleright$\ }{}
\SetCommentSty{mycommfont}

\usepackage{enumitem}
\usepackage{nicefrac}

\makeatletter
\@ifundefined{newcounteralias}{}{%
  \renewcommand\thmt@autorefsetup{%
    \@xa\def\csname\thmt@envname autorefname\@xa\endcsname\@xa{\thmt@thmname}%
  }%
}
\makeatother

\newtheorem{theorem}{Theorem}[section]
\newtheorem{lemma}[theorem]{Lemma}
\newtheorem{conjecture}[theorem]{Conjecture}
\newtheorem{corollary}[theorem]{Corollary}

\newtheorem{claim}[theorem]{Claim}
\newtheorem*{claim*}{Claim}
\newtheorem{definition}[theorem]{Definition}

\newtheorem{remark}[theorem]{Remark}

\allowdisplaybreaks

\declaretheoremstyle[
    bodyfont=\normalfont  %
]{upright}

\usepackage{graphicx}
\usepackage[colorlinks=true, allcolors=blue]{hyperref}
\usepackage[nameinlink]{cleveref}
\newcommand{\eps}{\varepsilon}

\newcommand{\congest}{\textsc{congest}\xspace}

\newcommand{\calF}{\mathcal{F}}

\newcommand{\calP}{\mathcal{P}}

\newcommand{\bw}{\mathbf{w}}
\newcommand{\bc}{\mathbf{c}}
\newcommand{\bx}{\mathbf{x}}
\newcommand{\by}{\mathbf{y}}
\newcommand{\bz}{\mathbf{z}}
\newcommand{\be}{\mathbf{e}}

\newcommand{\bq}{\mathbf{q}}
\newcommand{\bp}{\mathbf{p}}
\newcommand{\bv}{\mathbf{v}}
\newcommand{\br}{\mathbf{r}}
\newcommand{\ba}{\mathbf{a}}
\newcommand{\bb}{\mathbf{b}}
\newcommand{\blam}{\bm{\lambda}}

\newcommand{\odd}{\mathrm{odd}}
\newcommand{\even}{\mathrm{even}}
\newcommand{\lay}{\mathrm{layer}}

\DeclareMathOperator{\upd}{Upd}
\DeclareMathOperator{\dist}{dist}

\newcommand{\ZZ}{\mathbb{Z}}

\usepackage[colorinlistoftodos,textsize=tiny,textwidth=2cm,color=green!50!gray]{todonotes}

\newlength\hwhatcwidth%
\newlength\hwhatuwidth%
\newcommand*\halfwidehat[1]{%
  \settowidth\hwhatcwidth{\ensuremath{#1}}%
  \settowidth\hwhatuwidth{\ensuremath{\mathrm{#1}}}%
  \addtolength\hwhatuwidth{-\hwhatcwidth}%
  \makebox[0pt][l]{%
    \kern\dimexpr0.25\hwhatcwidth-0.667\hwhatuwidth\relax%
    \ensuremath{\widehat{\vphantom{#1}\rule{0.5\hwhatcwidth}{0pt}}}%
  }#1%
}%

\usepackage{todonotes}

\newcommand{\eat}[1]{}
\renewcommand{\emptyset}{\varnothing}

\crefname{observation}{observation}{observations}
\Crefname{observation}{Observation}{Observations}
\Crefname{proposition}{Proposition}{Propositions}
\Crefname{conjecture}{Conjecture}{Conjectures}
\Crefname{open}{Open Question}{Open Questions}

\Crefname{algocf}{Algorithm}{Algorithms}
\crefname{algocfline}{line}{lines}
\Crefname{invariant}{Invariant}{Invariants}
\Crefname{claim}{Claim}{Claims}
\Crefname{subclaim}{Subclaim}{Subclaims}
\Crefname{restatethm}{Theorem}{Theorems}
\Crefname{lem}{Lemma}{Lemmas}
\Crefname{restatelem}{Lemma}{Lemmas}

\newcommand{\diam}{\mathrm{diam}}

\allowdisplaybreaks

\title{Polynomial Lower Bounds for Steiner Shortcuts\\ via Test Functions}
\author{}
\author{Aaron Bernstein\thanks{New York University} \and Henry Fleischmann\thanks{Carnegie Mellon University} \and Bernhard Haeupler\thanks{INSAIT, Sofia University ``St.~Kliment Ohridski'' and ETH~Zurich; Partially funded by the Ministry of Education and Science of Bulgaria's support for INSAIT as part of the Bulgarian National Roadmap for Research Infrastructure, and by the European Research Council (ERC) under the European Union's Horizon Europe research and innovation programme (ERC Advanced grant agreement 101268062 and ERC Starting grant agreement 949272).} 
\and Gary Hoppenworth\thanks{TTIC} \and Thatchaphol Saranurak\thanks{University of Michigan}}
\date{}

\begin{document}
\maketitle
\pagenumbering{gobble}

\begin{abstract}

The \emph{Steiner shortcut} problem asks whether every directed graph $G=(V,E)$ with $n$ vertices and $m$ edges admits an augmentation $G_{+}=(V\cup V_{+},E\cup E_{+})$ with $m^{1+o(1)}$ edges such that $G_+$ has the same reachability structure as $G$ (i.e. for all $s,t\in V$, $s$ reaches $t$ in $G$ iff $s$ reaches $t$ in $G_+$), and the diameter of $G_{+}$ is $m^{o(1)}$. This is a core combinatorial question underlying many reachability algorithms in the dynamic, streaming, and CONGEST models, and \emph{all} parallel algorithms with almost-linear work and sublinear depth ~\cite{Fineman20,LiuJS19,CaoFR20,AshvinkumarBGS26}.

A sequence of works ruled out such a shortcut $G_{+}$ in the \emph{non-Steiner} setting where $V_{+}=\emptyset$ ~\cite{Hesse03,huang2021lower,LuWWX22,BH23,WilliamsXX24,HXX25}. However, as recently shown in \cite{bernstein2026reviving}, all known non-Steiner lower-bound constructions fail once Steiner vertices $V_{+}$ are allowed.

This paper rules out almost-optimal Steiner shortcuts: we show an infinite family of graphs $G=(V,E)$ with the property that, for any $\varepsilon>0$, a Steiner shortcut $G_{+}=(V\cup V_{+},E\cup E_{+})$ with $m^{1+\varepsilon}$ edges must have diameter at least $n^{1/56-O(\varepsilon)}$. 
% This establishes a definite limitation of the shortcut-based framework widely used in directed reachability and shortest-path algorithms.
This establishes a definitive barrier for the shortcut-based framework widely used in directed reachability and shortest-path algorithms, even in its most general form.

Our construction strengthens the standard non-Steiner lower-bound framework via an additional structure that allows Steiner vertices to be replaced by original vertices; we can then reduce back to the standard arguments without Steiner vertices. This new structure is achieved via a new general strategy based on \emph{test functions}, which is quite flexible and can potentially be extended to applications beyond our work. 

We also generalize our lower bound to exact Steiner hopsets in \emph{undirected} graphs and to depth reduction for monotone circuits. 
The latter significantly strengthens the classical lower bound based on Sylvester matrices \cite{JuknaSergeev2013}.
\end{abstract}

\newpage
\tableofcontents
\newpage 

\pagenumbering{arabic}
\section{Introduction}

Over the past few decades, shortcut sets and hopsets have emerged as widely used tools for computing reachability and shortest paths in a graph. The motivation is that in many models of computation, these problems are significantly easier in low-diameter graphs. Even if the input graph $G$ has large diameter, the diameter can be reduced by adding new edges to the graph. 
%The key idea behind shortcut sets is that even if the input graph $G = (V,E)$ has paths with many edges, these quantities can be reduced by adding new edges to $G$. 
Formally, a set of new edges $E_{+}$ is a $\beta$-\emph{shortcut set} if {\bf 1)} $G_+ = (V, E \cup E_+)$ has the same reachability structure as $G$; that is, for any pair $s,t \in V$, $s$ can reach $t$ in $G$ if and only if $s$ can reach $t$ in $G_+$, and {\bf 2)} $G_+$ has diameter at most $\beta$, where diameter is defined as the largest finite distance. A \emph{hopset} naturally generalizes this definition to capture shortest paths in addition to reachability. 

This simple idea leads to the following \emph{shortcut framework}. To compute reachability (resp. shortest paths) in an input graph $G$:
\begin{enumerate}
\item First compute a shortcut set (resp. hopset) $E_+$ of $G$, and 
\item Then compute reachability (resp. shortest paths) in $G_+ = (V,E \cup E_+)$
\end{enumerate}

This shortcut framework has been used extensively across many models of computation for both directed and undirected graphs, including parallel~\cite{ullman1990high,KleinS97,shi1999time,cohen2000polylog,MillerPVX15,ElkinN2016,elkin2019linear,LiuJS19,Fineman20,cao2020efficient,CaoFR20,AndoniSZ20,ElkinM21,karczmarz2021deterministic,CaoF23,AshvinkumarBCGH24,AshvinkumarBGS26}, dynamic~\cite{Bernstein09,Bernstein13,HenzingerKN2014STOC,HenzingerKN2014,Chechik18,GutenbergW20,BernsteinGW20,LackiN22}, streaming~\cite{HenzingerKN2016,ElkinN2016,Elkin20,ElkinT22}, \congest~\cite{Nanongkai14,HenzingerKN2016,ElkinN2016,ForsterN18,LiuJS19,Elkin20,CaoFR21,CaoF23,AshvinkumarBCGH24}, and others~\cite{CensorHillelDKL19,DinitzN19,ElkinN22,AshvinkumarBCGH24}.

One particularly notable application is the tantalizingly simple question of parallel reachability: given two vertices $s$ and $t$, how many parallel rounds are needed to determine whether there exists a path from $s$ to $t$? It is easy to achieve low depth via matrix multiplication, but if we restrict ourselves to algorithms with almost-linear work, then \emph{all} known algorithms with sublinear depth use the shortcut framework, both for directed reachability~\cite{Fineman20,LiuJS19,CaoFR20,AshvinkumarBGS26} and for directed shortest paths~\cite{cao2020efficient,CaoF23,AshvinkumarBCGH24,FischerHLRS25,AshvinkumarBGS26}. The state of the art achieves depth $n^{1/2+o(1)}$~\cite{LiuJS19}, and achieving depth $n^{o(1)}$ remains a central open question.

\paragraph{Existence of shortcut sets.}
The task of computing a shortcut set efficiently can itself be quite challenging, but putting this computational aspect aside, there is a fundamental limitation to the effectiveness of the framework: there needs to exist a good shortcut set! In particular, the ideal shortcut set (or hopset) $E_+$ should have $m^{1+o(1)}$ edges and yield hopbound $n^{o(1)}$. Such approximate hopsets indeed exist for undirected graphs~\cite{cohen2000polylog,ElkinN2016}, and they were a central ingredient in breakthrough algorithms for approximate shortest paths in parallel~\cite{cohen2000polylog}, dynamic~\cite{HenzingerKN2014}, distributed~\cite{HenzingerKN2016}, and streaming~\cite{HenzingerKN2016,ElkinN2016} models.

The search for ideal shortcut sets in directed graphs goes back to
Thorup~\cite{thorup1993shortcutting}, who conjectured that they exist and constructed them for planar
graphs~\cite{thorup1995shortcutting}, later extended to minor-free graphs~\cite{bhattacharyya2012transitive}.

For general directed graphs, unfortunately,
Hesse~\cite{Hesse03} 
showed that such ideal shortcut sets need not exist.
% Unfortunately, for directed graphs, \cite{Hesse03,huang2021lower} showed that some graphs do not admit this kind of ideal shortcut sets. 
A series of improvements~\cite{huang2021lower,LuWWX22,BH23,WilliamsXX24} led to the following lower bound:
\begin{theorem}[Adapted from \cite{HXX25}]
\label{thm:regular-shortcuts-lb}
There exists an infinite family of directed graphs $G = (V,E)$ with $|V| = n$ and $|E| = m$ such that, for any $\varepsilon > 0$, any shortcutting $G' = (V, E \cup E')$ with $O(m^{1+\varepsilon})$ edges has diameter $n^{2/9 - O(\varepsilon)}$.
\end{theorem}
These lower bounds had significant implications. They suggested that, although the shortcut framework has led to many polynomial improvements in directed graphs, it could never lead to almost-optimal algorithms.
% \hfnote{``truly fast'' should maybe be qualified} 
%In particular, it can never achieve depth below $n^{2/9}$ for parallel reachability with almost-linear work.

\paragraph{Steiner shortcuts.}
As pointed out by a recent paper \cite{bernstein2026reviving}, Theorem  \ref{thm:regular-shortcuts-lb} does not actually rule out the shortcut framework, because the framework does not require $G_+$ to be a strict shortcut set: $G_+$ can be any low-depth graph, as long as it preserves the reachability structure of the original $G$. To this end, the authors defined the notion of a \emph{Steiner} shortcut $G_+ = (V \cup V_+, E \cup E_+)$, which is any new graph that has the same reachability structure as $G$ for vertices in the original $V$.

From the perspective of the shortcut framework, a Steiner shortcut set is just as good as a regular shortcut set. But, as the authors of \cite{bernstein2026reviving} showed, all existing shortcut set lower bounds fail catastrophically once we allow Steiner vertices. This led them to hope for the following conjecture:

\begin{conjecture}[Rephrased from \cite{bernstein2026reviving}]
  \label{conjecture:steiner}
Every directed graph $G = (V,E)$ with $|V| = n$ and $|E| =m$ admits a Steiner shortcut $G_+= (V \cup V_+, E \cup E_+)$ with $|E_+| = m^{1 + o(1)}$ and diameter $n^{o(1)}$. \end{conjecture}

If true, the conjecture would have provided a promising direction for low-depth parallel reachability, as well as for many other problems. 

\subsection{Our Main Contribution: Disproving the Steiner Shortcut Conjecture}
Our main result disproves Conjecture \ref{conjecture:steiner}.

\begin{restatable}{theorem}{SteinerShortcutLowerBound}
\label{thm:intro-shortcut-lb}
There exists an infinite family of directed graphs $G=(V,E)$ with $|V|=n$ and $|E|=m$ such that, for any $\varepsilon>0$, any Steiner shortcut $G_+ = (V \cup V_+, E\cup E_+)$ with $|E_+|=O(m^{1+\varepsilon})$ edges has
$\diam(G_+) \ge n^{1/56 - O(\varepsilon)}$.
\end{restatable}

% \begin{restatable}{theorem}{SteinerShortcutLowerBound}
% \label{thm:intro-shortcut-lb}
% There exists an infinite family of graphs $G=(V,E)$ on $n$ nodes
% and $m$ edges such that any Steiner shortcut
% $G_+ = (V \cup V_+, E\cup E_+)$ with $|E_+|=O(m^{1+\varepsilon})$ edges has
% $\diam(G_+) \ge n^{1/56 - O(\varepsilon)}$.
% \end{restatable}

\paragraph{Implications.}
Theorem \ref{thm:intro-shortcut-lb} gives firm grounding for the claim that the shortcut framework cannot lead to almost-optimal algorithms. This is quite significant, because as mentioned above, this framework continues to be widely used. Most notably, it is the only known framework for low-depth reachability with almost-linear work, and our result shows that it cannot attain $n^{o(1)}$ depth.

% \aaroninline{Thatchaphol, you might want to rewrite the para below. Also add citations.} The only contending framework is based on interior point methods, which were recently used to obtain $\otil(\sqrt{n})$ depth and almost-linear work in moderately dense graphs~\cite{vanDenBrandGJV2025}. But interior-point methods are also have a fundamental structural barrier at $\sqrt{n}$ rounds, and even for relevant cases there is a long-standing barrier of $n^{1/3}$ CITE. We can thus conclude that the desired parallel reachability result will require a fundamentally new set of tools. 

The only contending framework is interior-point methods (IPMs), which recently achieved $\widetilde O(\sqrt {n})$ depth and $\widetilde O(m+n^{3/2})$ work for the more general min-cost flow problem~\cite{vanDenBrandGJV2025}. However, improving the $\sqrt n$ iteration bound is itself a major challenge in IPM theory. Standard self-concordant analysis gives this bound. Moreover, Todd and Ye proved that certain classes of long-step primal-dual IPMs require $\Omega(n^{1/3})$ iterations even to reduce the duality gap by a constant factor~\cite{ToddYe1996}.
Together with our shortcut lower bound, this suggests that almost-linear-work, $n^{o(1)}$-depth reachability requires substantially new ideas.

% \aaron{say something about it being interesting from a lower-bound perspective? I'm wierdly stuck on what to say here. And I'm not sure if it belongs here or in the short technical overview at the end of the intro.}

\paragraph{Open questions.}
While our lower bound rules out truly efficient Steiner shortcuts, our bound of $n^{1/56}$ is much smaller than the $n^{2/9}$ lower bound of~\cite{HXX25}. We prioritized having a clear, easy-to-understand refutation of the Steiner shortcut conjecture, so our exponent is not optimized. Nonetheless, even if pushed, our current techniques seem very far from achieving $n^{2/9}$. 

This gap points to two interesting research directions. Firstly, our lower bound still leaves open the possibility that Steiner shortcuts are more effective than regular shortcuts: can we construct Steiner shortcuts with $m^{1+o(1)}$ edges that have diameter $o(n^{2/9})$? Even constructions whose diameter is smaller than the best known \emph{upper} bounds for regular shortcuts (currently $n^{1/3}$ \cite{kogan2023faster}) would be highly interesting and have many potential algorithmic applications. On the other hand, it would also be interesting to develop more refined lower bound techniques that improve the exponent of our Theorem \ref{thm:intro-shortcut-lb}.
\begin{remark}
After extended interaction and being provided a very preliminary draft of the paper, ChatGPT 6 Astra claimed an optimized construction yielding a diameter lower bound of $n^{1/26}$ instead of $n^{1/56}$. Later, ChatGPT 6.1 (Sol) produced a Lean certificate, which the authors can share upon request. However, we are purposefully restricting our paper to the weaker Theorem \ref{thm:intro-shortcut-lb}, because the proof is much cleaner and more modular, and because it more effectively communicates the key new techniques.
\end{remark}

\subsection{Other Contributions}

\paragraph{Exact Steiner hopsets in undirected graphs.}
Hopsets play the same role for shortest paths that shortcut sets
play for reachability. An \emph{exact hopset} adds weighted edges so that every connected pair of original vertices has a shortest
path using few hops, while preserving all distances between
original vertices. 
% A \emph{Steiner hopset} may also introduce new vertices.

Again, known lower bounds for exact hopsets only work in
the non-Steiner setting~\cite{BH23,HXX25}.
We give the first lower bound ruling out almost-optimal exact Steiner hopsets, even for undirected, unweighted  graphs.
Specifically,
\Cref{cor:intro-hopset-lb} shows that any exact Steiner hopset with
$O(m^{1+\varepsilon})$ edges must have hopbound at least
$n^{1/56-O(\varepsilon)}$.
See \Cref{sec:hopset-ext} for details.

\paragraph{Depth reduction for monotone circuits.}
We also substantially strengthen the following classical depth reduction lower bound for OR-circuits.\footnote{These are circuits composed of only OR-gates, also known as \emph{rectifier circuits}, and were first introduced by Lupanov \cite{lupanov1956rectifier}. See the monograph of Jukna and Sergeev \cite{JuknaSergeev2013} for an excellent survey.} 
Sylvester matrices define functions
$f:\{0,1\}^n\to\{0,1\}^n$ that admit OR circuits with
$m=O(n\log n)$ wires, but for which, for every fixed $d$,
any OR circuit with depth $d$ requires $\Omega(n^{1+1/d})=\tilde{\Omega}(m^{1+1/d})$ wires.

In \Cref{cor:intro-monotone-circuit-lb}, we show an infinite family admitting OR circuits with $m$ wires, but for which any monotone circuit (containing both ANDs and ORs) with at most $n^{1/56-O(\varepsilon)}$ depth requires
$m^{1+\varepsilon}$ wires.
Thus, in the regime of $m^{1+\varepsilon}$
wires, we significantly improve the depth lower bound from constant to polynomial,  
even when AND gates are allowed.
See \Cref{sec:monotone-ckt-ext} for details.

We note that the connection between Steiner shortcuts and depth reduction for OR-circuits was previously observed in \cite{bernstein2026reviving}.

% Such OR-circuits are classical objects; they are also known as rectifier circuits and were first introduced by  Lupanov in 1956 \cite{lupanov1956rectifier}. A central question is their size under depth restrictions, a topic of a significant number of works \cite{Tarjan1975,ChandraFortuneLipton1983,ChandraFortuneLipton1985,Grinchuk1986,JuknaSergeev2013,IvanEtAl2014,Sergeev2022Coverings,AlmanLi2025}. In particular, see the monograph of Jukna and Sergeev \cite{JuknaSergeev2013} for an excellent survey and discussion of related problems. 

% \aaroninline{Someone else fill this out! Up to you whether you want to just have a short paragraph and leave a more details discussion for later, or whether we should have a discussion in the intro. I guess it depends on how well we can sell the problem.}

\subsection{Our Techniques}
% \aaroninline{An attempt to shorten the techniques section. I basically tried to get rid of everything that's not absolutely necessary for introducing the concept of test functions. It's still pretty long, since I don't know how to avoid the setup. Please go ahead and take a stab at shortening even further!}
% \aaroninline{I still think it works slightly better as a separate section, but also have no objection to integrating it with the pevious subsection.}

All existing lower bounds for non-Steiner shortcut sets use the same basic template: they construct a graph $G$ with a set $\mathcal{P}$ of many \emph{critical paths} such that every critical path is the unique path between its endpoints, and any two critical paths are nearly disjoint. But, as shown in \cite{bernstein2026reviving}, these properties only suffice to rule out non-Steiner shortcuts. We start by observing that, to rule out Steiner shortcuts, it is  sufficient to ensure a third property: any two \emph{compatible} critical paths $P_1, P_2 \in \calP$ overlap on a common vertex. Here compatibility means that the starting vertices of the paths can reach the endpoints of both paths.

Our lower bounds follow by constructing graphs with this overlap property. In fact, to ensure the overlap property, we end up proving properties about the \emph{cross-paths} of compatible critical paths. These are paths $P_{1,2}$ from the start of critical path $P_1$ to the end of another critical path $P_2$. Explicitly, we ensure that all such cross-paths $P_{1,2}$ are decomposable as a subpath of $P_1$ followed by a subpath of $P_2$. This then implies that $P_1$ and $P_2$ intersect at the crossing, and the existence of such a $P_{1,2}$ is ensured by compatibility.  However, reasoning about cross-paths requires fundamentally new tools; previous constructions only had to analyze the explicitly constructed critical paths, whereas cross-paths are a priori totally unstructured.

Our main technical innovation is a general framework for lifting properties of compatible critical paths to their cross-paths. We sketch this framework now.

At a high level, all Steiner/non-Steiner shortcut set lower bound constructions (including ours) represent the vertices as vectors, and also have a set of directions $C$. The edges are defined by an update function $\upd(v,c)$ which additively shifts a vertex to another vertex. The first lower bound constructions for (non-Steiner) shortcuts had simple linear update functions; the state-of-the-art bounds choose specific functions that are tailored to induce certain properties.

Rather than crafting a particular function $\upd(v,c)$, we introduce a general framework that allows us to take \emph{any} reasonable property that holds for the critical paths themselves, and then modify $\upd(v,c)$ in a black-box way to ensure that this property will also hold for all cross-paths; loosely speaking, the property must be expressible as a low-degree polynomial called a \emph{test function}. The property of cross-paths decomposing into the two corresponding critical paths can be expressed as a degree-four polynomial, hence defining a corresponding $\upd(v,c)$ and immediately resulting in a surprisingly simple refutation of \Cref{conjecture:steiner}.

 The flexibility of our approach significantly increases the expressivity of the general construction framework. For example, we show how to significantly optimize our simple lower bound construction under the same framework, just by using slightly more sophisticated cross-path properties and test function representations. We believe this general framework is likely to be useful for follow-up work on this problem.

We believe that test functions may also be useful for other lower bounds in directed graphs. Dense collections of long, unique paths are a basic ingredient in many such lower bounds~\cite{Hesse03,abboud2024reachability,BodwinHT23}, but these lower bounds often need additional properties as well. For example, lower bounds for the directed flow-cut gap also need high connectivity~\cite{ChuzhoyK09}, and lower bounds for DAG covers also need many pairs of paths that cannot lie in a common DAG~\cite{AssadiHW25}. Test functions give a generic way to add new properties to the \emph{cross-paths} of a collection of unique paths, as long as each property can be expressed by a low-degree polynomial. 
By enabling constructions with previously inaccessible restrictive properties, this new ingredient could lead to better lower bounds for existing problems such as the directed flow-cut gap, and for future lower bound problems in directed graphs.

\section{Steiner Shortcut Lower Bounds}
\label{sec:non-optimized}
% \aaron{rewrite this section to work with intro. I think we should state the non-optimized bound here, and say that all of Section 2 focuses on this non-optimized bound.}
In this section, we disprove the Steiner shortcut conjecture of \cite{bernstein2026reviving}. To better highlight the key insights, we prove the less optimized version below, and defer the proof of the more optimized bound of \Cref{thm:intro-shortcut-lb} to \Cref{sec:optimized}.
\begin{restatable}{theorem}{PolynomialSteinerShortcutLowerBound}
\label{thm:st-shortcut-lb}
There exist an absolute constant $\delta > 0$ and an infinite
family of (layered) DAGs $G=(V,E)$ on $n$ nodes and $m$ edges
such that any Steiner shortcutting $G_+=(V \cup V_+,E \cup E_+)$ either has
$|E_+|=\Omega(m^{1+\delta})$ or
$\diam(G_+)=\Omega(n^\delta)$.
\end{restatable}

We prove \Cref{thm:st-shortcut-lb} by showing a generic framework for augmenting existing (non-Steiner) shortcut set lower bounds into Steiner shortcut lower bounds. Roughly speaking, this amounts to taking the existing lower bound constructions based on average-free sets and augmenting each vertex with extra coordinates. The extra coordinates encode sufficient extra structure to rule out efficient Steiner shortcuts. Of course, these extra coordinates inevitably reduce the lengths of our critical paths, weakening the lower bounds from the non-Steiner setting, but the additional coordinates remain compact enough to still yield \Cref{thm:st-shortcut-lb}. 

\subsection{Sufficient properties for Steiner shortcut lower bounds} 
\label{sec:sufficient-props}
We begin with an abstract setup to motivate our design choices. This allows us to isolate the key properties exploited in our final explicit construction. As usual in the literature, our construction is a family of layered DAGs $G$ on $L$ total layers, with edges restricted to consecutive layers. Our analysis focuses on a particular family of \emph{critical paths} $\calP$ in $G$, where each path starts in the first layer and ends in the last layer. We show that Steiner shortcuttings cannot efficiently shortcut all the paths in $\calP$. 

\paragraph{Useful notation.}{
For a path $P$ in $G$, write $s(P)$ for the start of the path and $t(P)$ for its end. Additionally, for $v, w \in V$
, write $v \preceq w$ if $v$'s layer is at most $w$'s layer; write $v \prec w$ if $v$'s layer strictly precedes $w$'s. For any arbitrary graph $H$ and $u,v \in V(H)$, write $d_H(u,v)$ for the (hop)-distance between $u$ and $v$ in $H$.
}

We begin by describing sufficient properties of $\calP$ to show that, for any Steiner shortcutting of $G$, there exists a \emph{non-Steiner} shortcutting of $G$ on at most as many edges which shortcuts $\calP$ equally effectively. Since we construct our hard instances to be augmentations of existing non-Steiner shortcut lower bound constructions, they are themselves shortcut lower bound instances. Hence, such a reduction from Steiner shortcuttings to non-Steiner shortcuttings implies a Steiner shortcut lower bound.

The most basic question in studying Steiner shortcut sets is \emph{what do Steiner points accomplish}? In trying to measure this, suppose you are given a Steiner shortcutting $G_+$ of $G$. Assign each $P \in \calP$ a shortest path $P_+$ in $G_+$ between the same endpoints. Morally speaking, a Steiner point $v \in V_+$ is effective if it helps shortcut many paths in $\calP$; that is, $v$ belongs to many shortcutted paths $P_+$. But, since $G_+$ must preserve reachability between nodes in $V$, the existence of a Steiner point $v$ that belongs to many shortcutted paths is far from guaranteed. Indeed, the existence of such a Steiner point $v$ implies that the starts and ends of each critical path shortcutted via $v$ can all reach each other!

We isolate this structure in the notion of a \emph{compatible family} of critical paths. 
\begin{definition}[Compatible Families] \label{defn:compatible-family}
Let $\calF \subseteq \calP$. A family of critical paths $\calF$ is called a \emph{compatible family} if for all paths $P, Q \in \calF$, $s(P)$ can reach $t(Q)$ in $G$. A compatible family of size two is called a \emph{compatible pair}.
\end{definition}

Each Steiner point $v \in V_+$ corresponds to a compatible family: namely, the paths $P \in \calP$ whose shortcuttings $P_+$ pass through $v$. So in order to rule out efficient shortcuts, we need to construct $G$ so that its compatible families have restrictive structure. We show two simple properties of $\calP$ which ensure that Steiner shortcuttings are no stronger than non-Steiner shortcuttings.

\begin{restatable}[Steiner Shortcut Lower Bound Template] {restatethm}{lbtemplate}\label{thm:steiner-shortcut-template}
Let $G = (V,E)$ be a family of layered DAGs with critical paths $\calP$ each of length $D$ such that:
\begin{enumerate}[label = (P\arabic*)]
    \item (Critical paths are unique.) For each $P \in \calP$, $P$ is the unique path in $G$ between $s(P)$ and $t(P)$. \label{prop:unique}
    \item (Compatible pairs overlap.) For each compatible pair $P, Q \in \calP$ of critical paths, $P$ and $Q$ share a common node. \label{prop:compatible-pairs}
\end{enumerate}
Then, for any Steiner shortcutting $G_+ = (V \cup V_+, E \cup E_+)$ of $G$, there exists a non-Steiner shortcutting $G' = (V, E \cup E')$ such that:
\begin{enumerate}
    \item $|E'| \leq |E_+|$, and 
    \item for all $P \in \calP$, $d_{G'}(s(P), t(P)) \leq d_{G_+}(s(P), t(P))$.
\end{enumerate}
\end{restatable}
In fact \ref{prop:unique} and the property below are standard and suffice on their own to give non-Steiner shortcut lower bound constructions:
\begin{enumerate}[label = (P3)]
    \item (Critical paths are almost disjoint.) For each distinct pair $P, Q \in \calP$, the paths overlap on at most $\rho$ edges.  \label{prop:small-overlap}
\end{enumerate}
The other property, property \ref{prop:compatible-pairs}, is the new and interesting one. Constructions with all three properties are non-Steiner shortcut lower bounds from \ref{prop:unique} and \ref{prop:small-overlap}. Then, since \Cref{thm:steiner-shortcut-template} shows properties \ref{prop:unique} and \ref{prop:compatible-pairs} reduce Steiner shortcuts to non-Steiner shortcuts, constructions with all three properties are also Steiner shortcut lower bounds. Our eventual construction satisfies all three properties. 

For the rest of this subsection, we sketch the proof of \Cref{thm:steiner-shortcut-template}. The complete argument is deferred to \S\ref{sec:deferred-proof}. We begin by observing that compatible pairs sharing common nodes \ref{prop:compatible-pairs}, combined with uniqueness of paths \ref{prop:unique}, implies the much stronger property that every compatible family has a single common vertex, even if the family contains many paths.
\begin{restatable}{restatelem}{compatcommon}
\label{lem:compatible-families-common}
Properties \ref{prop:unique} and \ref{prop:compatible-pairs} imply that for every non-empty compatible family $\calF \subseteq \calP$, there exists an original vertex $v \in V$ that is contained in all paths of $\calF$.
\end{restatable}
\begin{proof}
First, observe that every \emph{subpath} of a critical path is unique between its
endpoints: otherwise, replacing that subpath would contradict
property~\ref{prop:unique}. In particular, the common vertices of two
critical paths form an interval of layers.

Now fix $P_0\in\calF$, and, for
each $P\in\calF$, let $I_P$ be the interval of layers on which $P$
and $P_0$ agree. Property~\ref{prop:compatible-pairs} says that these
intervals are nonempty. We claim that every pair of intervals intersects.
Indeed, take $P,Q\in\calF \setminus \{P_0\}$ and choose vertices
\[
 x\in P\cap P_0,\qquad y\in Q\cap P_0,\qquad z\in P\cap Q.
\]
Exchange $P,Q$ if necessary so that $x \preceq y$.
There are three cases, according to the layer of $z$ relative to the layers of $x$ and $y$; see \Cref{fig:compatible-family-cases}.
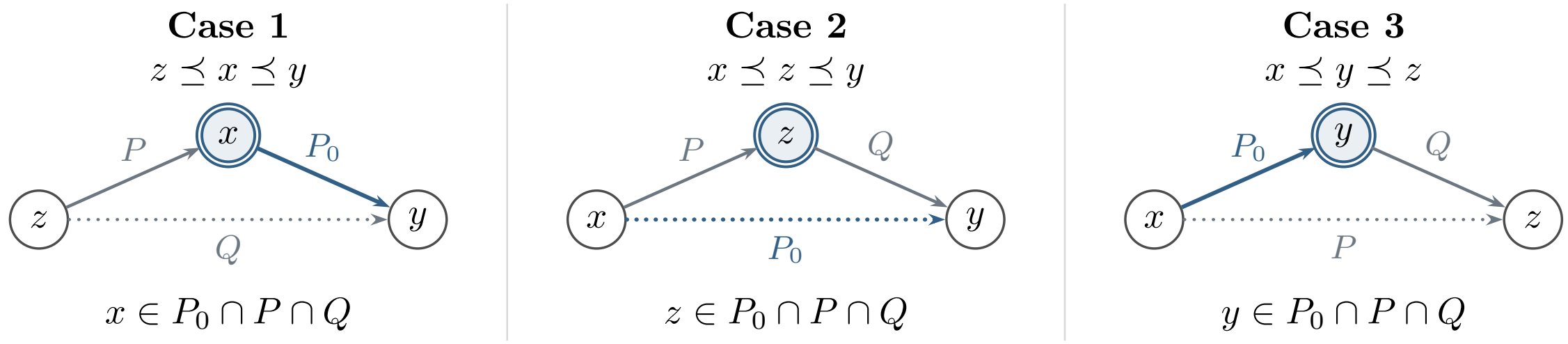
\begin{figure}[tbp]
    \centering
    \begingroup
\definecolor{pathblue}{HTML}{315F85}
\definecolor{edgegray}{HTML}{6C7782}
\begin{tikzpicture}[
    font=\small,
    >={Stealth[length=1.6mm,width=1.1mm]},
    vertex/.style={circle,draw=black!70,fill=white,line width=.65pt,
        minimum size=5.5mm,inner sep=0pt},
    common vertex/.style={vertex,draw=pathblue,fill=pathblue!10,
        double,double distance=.6pt,line width=.75pt},
    base path/.style={->,draw=pathblue,line width=1.1pt},
    other path/.style={->,draw=edgegray,line width=.85pt},
    comparison path/.style={dash pattern=on 0pt off 2.5pt,line cap=round},
    path label/.style={font=\footnotesize,inner sep=1pt,fill=white},
]

% Case 1: Q[z,y] = P[z,x] followed by P_0[x,y].
\begin{scope}
    \node[font=\small\bfseries] at (2.2,1.85) {Case 1};
    \node at (2.2,1.4) {$z\preceq x\preceq y$};
    \node[vertex] (z) at (.4,0) {$z$};
    \node[common vertex] (x) at (2.2,.8) {$x$};
    \node[vertex] (y) at (4,0) {$y$};

    \draw[other path] (z) --
        node[path label,above=3pt,text=edgegray] {$P$} (x);
    \draw[base path] (x) --
        node[path label,above=3pt,text=pathblue] {$P_0$} (y);
    \draw[other path,comparison path] (z) --
        node[path label,below=3pt,text=edgegray] {$Q$} (y);

    \node at (2.2,-.9) {$x\in P_0\cap P\cap Q$};
\end{scope}

\draw[black!15,line width=.5pt] (4.85,-1.15) -- (4.85,2.05);

% Case 2: P_0[x,y] = P[x,z] followed by Q[z,y].
\begin{scope}[xshift=5.3cm]
    \node[font=\small\bfseries] at (2.2,1.85) {Case 2};
    \node at (2.2,1.4) {$x\preceq z\preceq y$};
    \node[vertex] (x) at (.4,0) {$x$};
    \node[common vertex] (z) at (2.2,.8) {$z$};
    \node[vertex] (y) at (4,0) {$y$};

    \draw[other path] (x) --
        node[path label,above=3pt,text=edgegray] {$P$} (z);
    \draw[other path] (z) --
        node[path label,above=3pt,text=edgegray] {$Q$} (y);
    \draw[base path,comparison path] (x) --
        node[path label,below=3pt,text=pathblue] {$P_0$} (y);

    \node at (2.2,-.9) {$z\in P_0\cap P\cap Q$};
\end{scope}

\draw[black!15,line width=.5pt] (10.15,-1.15) -- (10.15,2.05);

% Case 3: P[x,z] = P_0[x,y] followed by Q[y,z].
\begin{scope}[xshift=10.6cm]
    \node[font=\small\bfseries] at (2.2,1.85) {Case 3};
    \node at (2.2,1.4) {$x\preceq y\preceq z$};
    \node[vertex] (x) at (.4,0) {$x$};
    \node[common vertex] (y) at (2.2,.8) {$y$};
    \node[vertex] (z) at (4,0) {$z$};

    \draw[base path] (x) --
        node[path label,above=3pt,text=pathblue] {$P_0$} (y);
    \draw[other path] (y) --
        node[path label,above=3pt,text=edgegray] {$Q$} (z);
    \draw[other path,comparison path] (x) --
        node[path label,below=3pt,text=edgegray] {$P$} (z);

    \node at (2.2,-.9) {$y\in P_0\cap P\cap Q$};
\end{scope}
\end{tikzpicture}
\endgroup
    \caption{The three layer orders in the proof of \Cref{lem:compatible-families-common}. In each case, uniqueness of subpaths of critical paths forces the middle vertex to lie on all three critical paths.}
    \label{fig:compatible-family-cases}
\end{figure}
\begin{enumerate}[label = ]
    \item \textbf{Case 1:} $z \preceq x \preceq y$. In this case, 
    $P[z,x]\circ P_0[x,y]$ is a path from $z$ to $y$.
    Uniqueness of $Q[z,y]$ forces this path to equal $Q[z,y]$,
    so $x$ belongs to all three paths.
    \item \textbf{Case 2: $x \preceq z \preceq y$.} In this case, uniqueness implies $P[x,z]\circ Q[z,y]$ must equal $P_0[x,y]$,
    so $z$ belongs to all three paths. 
    \item\textbf{Case 3:} $x \preceq y \preceq z$. Finally, in this last case
    $P_0[x,y]\circ Q[y,z]$ must equal $P[x,z]$.
    Thus $y$ belongs to all three paths.
\end{enumerate}
    
In every case, $I_P\cap I_Q$ is nonempty. Finally, finitely many
pairwise-intersecting intervals of layers have a common intersection: their largest
left endpoint is no larger than any of their right endpoints.
At that layer, every member of $\calF$ agrees with $P_0$. This gives a common vertex $v \in P_0 \subseteq V$.
\end{proof}

We remark that any family of critical paths $\calF \subseteq \calP$ with a common vertex is clearly compatible, so \Cref{lem:compatible-families-common} shows that, after assuming \ref{prop:unique} and \ref{prop:compatible-pairs}, compatibility is equivalent to having a common vertex.

Given that each Steiner vertex corresponds to a compatible family and each compatible family has a common vertex in the original graph, a natural strategy emerges. We can try to replace each Steiner point with the common vertex of its compatible family.
\begin{proof}[Sketch of the proof of \Cref{thm:steiner-shortcut-template}]
Let $G_+ = (V \cup V_+, E \cup E_+)$. For each critical path $P \in \calP$, let $P_+$ be a shortest path in $G_+$ between the same endpoints. Then the compatible family of critical paths corresponding to $x \in V_+$ is the set $\calF_x$ of $P \in \calP$ such that $P_+$ passes through $x$. 

By \Cref{lem:compatible-families-common}, for each $x \in V_+$, there exists $v_x \in V$ such that $v_x$ lies in all paths in $\calF_x$. We can imagine constructing $G'$ from $G_+$ by simply merging each $x \in V_+$ with the corresponding $v_x \in V$. See \Cref{fig:steiner-vertex-merge}.
\begin{figure}[tbp]
    \centering
    \begingroup
\definecolor{componentblue}{HTML}{315F85}
\definecolor{componentgray}{HTML}{6C7782}
\begin{tikzpicture}[
    font=\small,
    >={Stealth[length=1.7mm,width=1.2mm]},
    neighbor/.style={circle,draw=componentblue,fill=white,
        line width=.85pt,minimum size=6mm,inner sep=0pt},
    common vertex/.style={neighbor,fill=componentblue!8,
        double,double distance=.6pt,minimum size=7mm},
    steiner/.style={rectangle,rounded corners=1.3pt,
        draw=componentblue,fill=componentblue!7,line width=.85pt,
        minimum size=6mm,inner sep=0pt},
    shortcut/.style={->,draw=componentblue,line width=1.05pt,
        preaction={draw=white,line width=3.4pt,-,
            shorten <=1pt,shorten >=1pt}},
]

% Keep the neighbors and the chosen common vertex fixed in both panels.
\foreach \shift in {0,8.2} {
    \begin{scope}[xshift=\shift cm]
        \node[neighbor] (a) at (0,.85) {$a$};
        \node[neighbor] (b) at (0,-.85) {$b$};
        \node[common vertex] (vx) at (3.15,0) {$v_x$};
        \node[neighbor] (c) at (6.3,.85) {$c$};
        \node[neighbor] (d) at (6.3,-.85) {$d$};

        \ifdim\shift pt=0pt
            \node[font=\normalsize] at (3.15,2.55) {Steiner shortcutting $G_+$};
            \node[steiner] (x) at (3.15,1.65) {$x$};
            \draw[shortcut] (a) -- (x);
            \draw[shortcut] (b) -- (x);
            \draw[shortcut] (x) -- (c);
            \draw[shortcut] (x) -- (d);
        \else
            \node[font=\normalsize] at (3.15,2.55) {After merging $x$ with $v_x$};
            \draw[shortcut] (a) -- (vx);
            \draw[shortcut] (b) -- (vx);
            \draw[shortcut] (vx) -- (c);
            \draw[shortcut] (vx) -- (d);
        \fi
    \end{scope}
}

\draw[->,draw=componentgray,line width=.9pt,
    >={Stealth[length=2mm,width=1.5mm]}] (6.85,.1) -- (7.65,.1);
\node[font=\footnotesize] at (7.25,.65) {$x\mapsto v_x$};

\node[neighbor,minimum size=3.5mm] at (3.8,-1.65) {};
\node[anchor=west,font=\footnotesize] at (4.05,-1.65) {neighbor};
\node[steiner,minimum size=3.5mm] at (6.4,-1.65) {};
\node[anchor=west,font=\footnotesize] at (6.65,-1.65) {Steiner vertex};
\draw[shortcut] (9.4,-1.65) -- (10,-1.65);
\node[anchor=west,font=\footnotesize] at (10.2,-1.65) {shortcut};
\end{tikzpicture}
\endgroup
    \caption{Replacement of a Steiner vertex $x$ by a common vertex $v_x\in V$ of the paths in $\calF_x$. Edges incident to $x$ are redirected to $v_x$.}

    \label{fig:steiner-vertex-merge}
\end{figure}
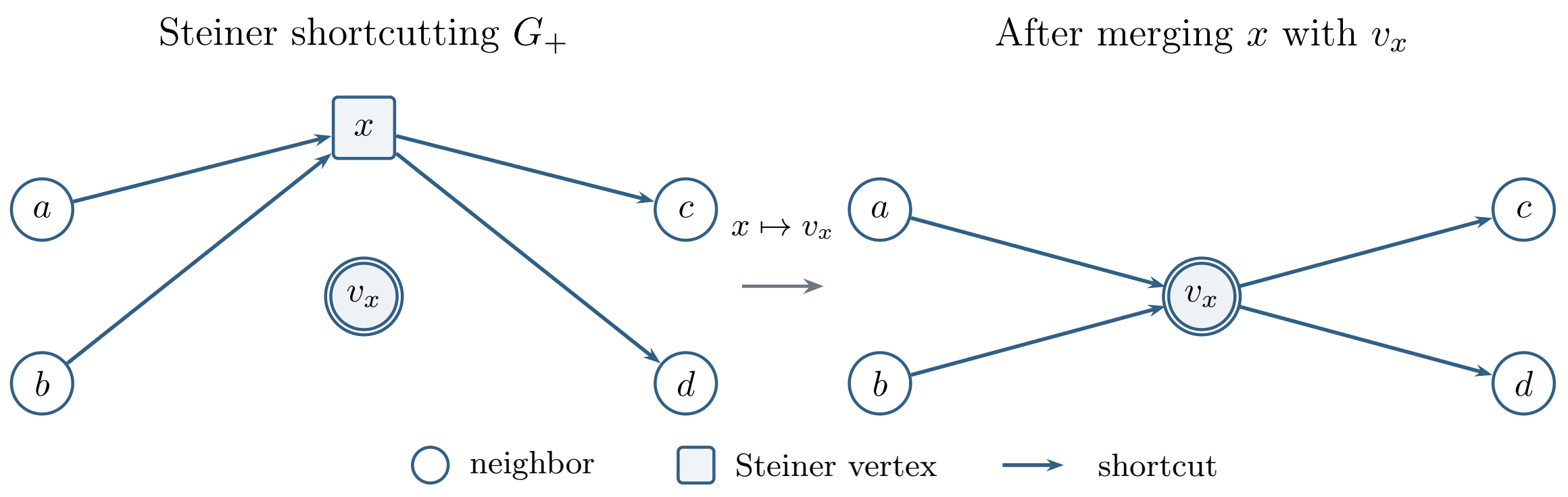
 Clearly this results in $|E'| \leq |E_+|$. Moreover, the paths $P_+$ are mapped via the merge to paths $P'$ in $G'$ of the same length. The trickier detail is to check whether the resulting graph $G'$ is actually a valid shortcutting; i.e., it preserves reachability of $G$. This amounts to checking that the layer of $v_x$ is between the layers of the vertices in $V$ that can reach and be reached by it. 

 When there is only one candidate for each $v_x$, this can be shown using the uniqueness of critical paths \ref{prop:unique}. However, when there are multiple candidates for each $v_x$, validity of this transformation actually ends up crucially depending on each choice.  We defer the details of proving the validity of $G'$ to the full proof of \Cref{thm:steiner-shortcut-template} in \S\ref{sec:deferred-proof}. 
  \renewcommand{\qedsymbol}{\ensuremath{\diamond}}
\end{proof}

Via \Cref{thm:steiner-shortcut-template}, we have reduced away all the power of Steiner points, just using properties \ref{prop:unique} and \ref{prop:compatible-pairs}. The remaining property \ref{prop:small-overlap} is essential to then proving a shortcut set lower bound, but this is standard, so we defer this detail for now.

\subsection{How to restrict compatible families}
The game now is to ensure \ref{prop:compatible-pairs}. This will determine how exactly we augment the coordinates of our shortcut set lower bound construction. 

\paragraph{The construction in more detail.} We begin by giving a bit more detail about our construction. Let $C \subseteq \ZZ^d$ be an \emph{average-free} set of critical directions. That is, for every 
positive integer $k$ and every $\bc_0, \bc_1,\ldots,\bc_k\in C$,
\begin{equation}\label{eq:average}
 k\bc_0=\bc_1+\cdots+\bc_k
 \quad\Longrightarrow\quad \bc_1=\cdots=\bc_k=\bc_0.
\end{equation} 
Our hard instance $G$ has some underlying geometry based on $C$; each vertex $v \in V$ is a high-dimensional vector $\bv = (\bp, \bq) \in \ZZ^N$, where $\bp = (\ell, \br, \bx)$ is the \emph{position} and $\bq$ is an augmentation we describe later. The coordinates of $\bp$ are:
\begin{itemize}
    \item $\ell \in \ZZ$, the layer number;
    \item $\br = (r_{\odd}, r_{\even}) \in \ZZ^2$, counters of the steps taken in odd/even directions, up to some initial offsets; and 
    \item $\bx = (\bx^{\odd}, \bx^{\even})\in \ZZ^{d + d}$, sums of the steps taken in odd/even directions, again up to some initial offsets.
\end{itemize}
In total, $\bp \in \ZZ^{2d + 3}$. Recall that $G$ has $L$ total layers. The edges in $G$ are induced by the following update function. For $\bp \in \ZZ^{2d+3}$ and $\bc \in C$, we define 
\begin{equation}
    \upd(\bp, \bc) := \begin{cases} (1, 1, 0, \bc, \mathbf{0}, \upd_{\bq}(\bp, \bc)) & \text{if } \bp[\ell] \equiv 1 \pmod{2} \\
    (1, 0, 1, \mathbf{0}, \bc, \upd_{\bq}(\bp, \bc))
    & \text{if } \bp[\ell] \equiv 0 \pmod{2}
    \end{cases}.
    \label{eq:update}
\end{equation}
Here $\upd_{\bq}(\bp, \bc)$ is a thus far unspecified function of $\bc$ and $\bp$. Then, for $\bv \in V$ with $\lay(\bv) < L$, $\bv$ has an edge to $\bw = \bv + \upd(\bv[\bp], \bc)$ for each $\bc \in C$. In words, the update to the position $\bp$ is simple and standard: increase the layer by $1$, increase the relevant odd/even counter by $1$, and increase the odd/even sum $\bx$ by $\bc$.

Our critical paths $\calP$ are defined by a vertex $\bv \in V$ with $\lay(\bv) = 1$ and \emph{critical directions} $(\ba,\bb) \in C^2$. The corresponding critical path $P$ is then given by $\bv^{(1)} = \bv$ and, for $i < L$,
\begin{equation*}
\bv^{(i+ 1)} = \begin{cases}
    \bv^{(i)} + \upd(\bv^{(i)}[\bp], \ba) &\text{if } i \equiv 1 \pmod{2} \\
  \bv^{(i)} + \upd(\bv^{(i)}[\bp], \bb) &\text{if } i \equiv 0 \pmod{2}
\end{cases}.
\end{equation*}
In words, this is the path from $\bv$ to the last layer where the edges alternate between taking the edge corresponding to $\ba$ and the edge corresponding to $\bb$. 

%By defining critical paths in this way, average-freeness of $C$ gives both properties \ref{prop:unique} and \ref{prop:small-overlap}. This consequence is standard, so we defer this detail for now.

\paragraph{Comparison to prior work.}
Without the augmentation $\bq$ and $\upd_{\bq}$, the above construction corresponds to the standard template for (non-Steiner) shortcut lower bounds. In particular, by defining critical paths as above, average-freeness of $C$ gives both properties \ref{prop:unique} and \ref{prop:small-overlap} (see e.g. \cite{huang2021lower,HXX25}); this consequence is standard, so we defer this detail for now. Our key innovation is in using $\bq$ to also ensure \ref{prop:compatible-pairs}.

\paragraph{The augmentation $\bq$.}
The idea for ensuring \ref{prop:compatible-pairs} is to restrict compatible families by showing that with enough extra information added to $\bq$, one can transfer properties from any pair of compatible critical paths onto their cross-paths.

Concretely, let $P_1, P_2 \in \calP$ be a compatible pair with critical directions $(\ba^{(1)}, \bb^{(1)})$ and $(\ba^{(2)}, \bb^{(2)})$, respectively. Choose arbitrary cross-paths
$P_{1,2}$ and $P_{2,1}$  from $s(P_1)$
to $t(P_2)$ and $s(P_2)$ to $t(P_1)$, respectively. We aim to ensure the following ``splicing property'' of cross-paths:
\begin{enumerate} [label = (P2$'$)]
    \item The cross-path $P_{1,2}$ can be decomposed as $P_1[s(P_1), v] \circ P_2[v, t(P_2)]$. \label{prop:splicing-prop}
\end{enumerate} 
In particular, \ref{prop:splicing-prop} implies that $P_1$ and $P_2$ have a common intersection at $v$ and hence implies \ref{prop:compatible-pairs}.

In our construction, the position vector $\bp$ is tailored to restrict the structure of the critical paths $P_i$, but tells us little about the cross-paths. We will show that by adding coordinates to $\bq$, we can restrict the set of valid cross-paths so that they must all satisfy \ref{prop:splicing-prop}. Observe, as a starting point, that the decomposition of \ref{prop:splicing-prop} trivially holds for the paths $P_1$, $P_2$ themselves. We will show how to use $\bq$ to ensure that a property that holds for any critical paths $P_1, P_2$ also holds for \emph{all} cross-paths $P_{1,2}$ and $P_{2,1}$. This technique is quite general, and is not restricted to \ref{prop:splicing-prop}; loosely speaking, we will show that any property that can be captured with a \emph{test function} that is a low-degree polynomial can be transferred from critical paths to all cross-paths. We now present the technique in its full generality: for the proof of \Cref{thm:st-shortcut-lb} we only apply the technique to \ref{prop:splicing-prop}, but in later optimizations (\Cref{sec:optimized}) we will apply it to other properties as well.

%Note that if we ignore $\bq$, the template above is tailored to restrict the structure of each critical path $P_i$, but does nothing to restrict the structure of the cross paths. Thus, if we look at existing realizations of this template (without $\bq$) (e.g. \abnote{add REF}, they can have a huge number of different crosspaths $P_{1,2}$, some of which will not satisfy \ref{prop:splicing-prop}. By adding additional coordinates to $\bq$, we are able to restrict 

%Observe that the decomposition in \ref{prop:splicing-prop} trivially holds for the paths $P_1$, $P_2$ themselves. But there can be many crosspaths $P_{1,2}$, and if we didn't have an augmentation $q$, 

%To show \ref{prop:splicing-prop}, we apply a general purpose technique for constructing $\bq$ to transfer properties of pairs critical paths to their cross-paths.
\paragraph{Test Functions}
For every property that we want to transfer from $P_1$ and $P_2$ to
$P_{1,2}$ and $P_{2,1}$, we define a \emph{test function}
$F_{P_1,P_2}(\bp,\bc)$, depending on the current position $\bp$ and the
step $\bc$, and with the following properties:
\begin{enumerate}[label = (T\arabic*)]
    \item $F_{P_1,P_2}(\bp,\bc) \geq 0$ for all $\bp \in \ZZ^{2d + 3}$ and $\bc \in C$. \label{prop:nonneg}
    \item If for each step $(\bv, \bc)$ of a path $P$ we have $F_{P_1,P_2}(\bv[\bp],\bc)=0$,
    then $P$ has the desired property.  In particular, we
    require $F_{P_1,P_2}$ to vanish at every step of $P_1$
    and $P_2$. \label{prop:vanishing} 
    \item $F_{P_1,P_2}(\bp,\bc)$ is a linear function of
    $\upd_{\bq}(\bp, \bc)$; that is, there exists a coefficient vector
    $\blam_{P_1,P_2}$ such that
    \[
        F_{P_1,P_2}(\bp,\bc)
        = \upd_{\bq}(\bp,\bc)^{\top}\blam_{P_1,P_2}.
    \]
    \label{prop:linear}
\end{enumerate}
Zooming out, test functions are the constraints on the update functions $\upd_{\bq}$ (as apparent in \ref{prop:linear}) which define the edges of our final graph. The goal is to design specific test functions that ensure \ref{prop:splicing-prop}. For now, let us first see why the above general properties of test functions are convenient. 

First, since $\bv[\bq]$ is updated by adding $\upd_{\bq}(\bp, \bc)$ at each
step along a path, its sum telescopes. That is, for a path $P = (\bv^{(1)}, \bv^{(2)}, \ldots, \bv^{(k)})$, we have 
\begin{equation}
     \sum_{i = 1}^{k-1} (\bv^{(i+1)}[\bq] - \bv^{(i)}[\bq]) =  \sum_{(\bv^{(i)},\bc^{(i)})\text{ along }P}\upd_{\bq}(\bv^{(i)}[\bp],\bc^{(i)})
    = t(P)[\bq] - s(P)[\bq]. \label{eq:use-of-linearity}   
\end{equation}
Crucially, this sum depends only on the endpoints of the path.

By the linearity property \ref{prop:linear}, every test function telescopes as well.
Write $F=F_{P_1,P_2}$ and $\blam =\blam_{P_1,P_2}.$
For any cross-paths $P_{1,2}$ and $P_{2,1}$, we obtain
\begin{align*}
\sum_{(\bp,\bc)\text{ along }P_1}F(\bp,\bc)
    +
    \sum_{(\bp,\bc)\text{ along }P_2}F(\bp,\bc) &= (t(P_1)[\bq]-s(P_1)[\bq]+t(P_2)[\bq]-s(P_2)[\bq])^\top \blam
    \\
    &=
    \sum_{(\bp,\bc)\text{ along }P_{1,2}}F(\bp,\bc)
    +
    \sum_{(\bp,\bc)\text{ along }P_{2,1}}F(\bp,\bc).    
\end{align*}
By property \ref{prop:vanishing}, the sums along $P_1$ and $P_2$ vanish.
Consequently,
\begin{equation}
   0
    =
    \sum_{(\bp,\bc)\text{ along }P_{1,2}}F(\bp,\bc)
    +
    \sum_{(\bp,\bc)\text{ along }P_{2,1}}F(\bp,\bc).  \label{eq:rhs}
\end{equation} 
Finally, by the non-negativity property \ref{prop:nonneg}, every summand on the right-hand side of \Cref{eq:rhs} must be $0$. Hence, $F$ vanishes at every step
of both cross-paths. Property \ref{prop:vanishing} then implies that the desired
property holds for $P_{1,2}$ and $P_{2,1}.$ E.g., in our application in proving \Cref{thm:st-shortcut-lb}, we construct a test function such that this yields \ref{prop:splicing-prop} which in turn gives \ref{prop:compatible-pairs}.

Notice that we crucially exploit the fact that $\bq$ is updated by adding $\upd_{\bq}(\bp, \bc)$ at each step; see \Cref{eq:use-of-linearity}. This is how we take advantage of the linearity property \ref{prop:linear}.

 To implement this test function technique, we need to somehow ensure that $F_{P_1,P_2}(\bp, \bc)$ is a linear function of $\upd_{\bq}(\bp, \bc)$. To do this, one could introduce a separate coordinate of $\bq$ for every test
function and set $\upd_{\bq}(\bp, \bc)[F_{P_1,P_2}] = F_{P_1,P_2}(\bp, \bc)$. This would trivially ensure that $F_{P_1,P_2}(\bp, \bc)$ is linear in $\upd_{\bq}(\bp, \bc)$. However, there are too many distinct pairs of critical paths for this approach to be viable. 
Instead, let $\mathcal{M}_\Delta$ be the set of all monomials
in the entries of $\bp$ and $\bc$ of total degree at most
$\Delta$, including the constant monomial $1$. We give $\bq$
one coordinate $\bq[z]$ for each $z \in \mathcal{M}_\Delta$
and define
\[
    \upd_{\bq}(\bp,\bc)[z] := z(\bp,\bc).
\]
Thus, each step adds $z(\bp,\bc)$ to the coordinate $\bq[z]$.
Every test function $F_{P_1,P_2}(\bp,\bc)$ that is a polynomial
of total degree at most $\Delta$ is a linear combination of
these monomials, and hence is a linear function of
$\upd_{\bq}(\bp,\bc)$.

For the test functions we construct in this section, $\Delta = 4$ suffices. (This bound is unoptimized to simplify the presentation.)
The same framework also accommodates more sophisticated,
higher-degree test functions, which may depend on more than
two paths.

In summary, our goal is to construct nonnegative test functions
$F_{P_1,P_2}(\bp,\bc)$ that are low-degree polynomials in $\bp,\bc$,
vanish along $P_1$ and $P_2$, and whose vanishing encodes a
desired property. The framework then transfers that property
to every pair of cross-paths $P_{1,2}$ and $P_{2,1}$.

The test functions below have polynomial expressions that
depend on the parity of the layer. To accommodate this, we
give $\bq$ two coordinates $\bq[z,\odd]$ and $\bq[z,\even]$
for each $z \in \mathcal{M}_\Delta$. For
$\sigma \in \{\odd,\even\}$, define
\[
    \upd_{\bq}(\bp,\bc)[z,\sigma] :=
    \begin{cases}
        z(\bp,\bc) & \text{if the layer $\bp[\ell]$ has parity $\sigma$},\\
        0 & \text{otherwise}.
    \end{cases}
\]
Any test function given by a polynomial of total degree at
most $\Delta$ on odd layers and another such polynomial on
even layers is then a linear function of
$\upd_{\bq}(\bp,\bc)$, ensuring \ref{prop:linear}.

\paragraph{Test functions for splicing.}
Consider the test function 
\begin{equation*}
R_{P_1, P_2}(\bp, \bc) = \beta(\bp, \bc) \alpha(\bp, \bc),    
\end{equation*}
where $\alpha$ is defined as follows:
\begin{equation*}
    \alpha(\bp, \bc) := \begin{cases}
        \| \bc - \ba^{(2)} \|_2^2 &\text{if } \bp[\ell] \equiv 1 \pmod{2} \\
         \| \bc - \bb^{(2)} \|_2^2 &\text{if } \bp[\ell] \equiv 0 \pmod{2}
    \end{cases}.
\end{equation*}
The function $\alpha$ is zero if and only if a step is in accordance with the critical directions of $P_2$. To define $\beta$, it is convenient to write $\bp' := \bp + \upd(\bp, \bc)[\bp] - s(P_1)[\bp]$. That is, $\bp'$ is the position $\bp$ after the update, offset by the starting position of $P_1$. Then, 
\begin{align*}
\beta(\bp, \bc) := \|\bp'[\bx^{\odd}] -  \bp'[r_{\odd}]\ba^{(1)} \|_2^2 + \|\bp'[\bx^{\even}]  - \bp'[r_{\even}] \bb^{(1)} \|_2^2.
\end{align*}
The function $\beta$ being zero tracks whether a path starting at $s(P_1)$ has thus far remained on $P_1$.  The functions $\alpha$ and $\beta$ are both clearly non-negative, and so $R_{P_1,P_2}$ is also non-negative. Moreover, since $P_2$ only takes steps in directions $\ba^{(2)}$ and $\bb^{(2)}$, the function $\alpha$, and hence $R_{P_1, P_2}$, vanishes on $P_2$. Similarly, since $P_1$ starts at $s(P_1)$ and updates:
\begin{itemize}
    \item the $\bx^{\odd}$ coordinate by $\ba^{(1)}$ and the $r_{\odd}$ coordinate by $1$ on odd layers, and
    \item the $\bx^{\even}$ coordinate by $\bb^{(1)}$ and the $r_{\even}$ coordinate by $1$ on even layers,
\end{itemize}
we have that $\beta$ vanishes along $P_1$ and hence so does $R_{P_1, P_2}$. Expanding the definitions of $\alpha$ and $\beta$, we see that $\alpha$ is a quadratic in $\bc$, and $\beta$ is quadratic in $\bp$ and $\bc$, barring the issue of the function depending on the parity of the layer. Hence, the product is a degree-four polynomial in $\bp$ and $\bc$ and linear in $\upd_{\bq}(\bp, \bc)$ as long as we record all monomials of total degree at most four in $\bp$ and $\bc$ in $\upd_{\bq}(\bp, \bc)$ (and duplicate coordinates for parity concerns).

Thus, the test function satisfies conditions \ref{prop:nonneg}, \ref{prop:vanishing}, and \ref{prop:linear} of our framework. Consequently, $R_{P_1, P_2}$ sums to $0$ along the cross-path $P_{1,2}$. Hence, either $\alpha$ or $\beta$ must be zero along every step of $P_{1,2}$. We show that this implies \ref{prop:splicing-prop} in the following claim.
\begin{claim} \label{cla:order-prop}
$R_{P_1, P_2}$ vanishing along $P_{1,2}$ for all distinct compatible $P_1, P_2 \in \calP$ and cross-paths $P_{1,2}$ implies \ref{prop:splicing-prop}.
\end{claim}
\begin{proof}[Proof of \Cref{cla:order-prop}]
First, observe that, along a path starting at $s(P_1)$, $\beta(\bp, \bc)$ is zero if and only if all steps thus far along the path (including the current $\bc$) have been along $P_1$. Indeed, otherwise, suppose there has been an odd step not in direction $\ba^{(1)}$ (the even step case is analogous). Then, thereafter, $\beta(\bp, \bc) \geq \| k \ba^{(1)} - \sum_{i=1}^k \bc_i  \|_2^2$ for $k \geq 1$ and $\bc_1, \dots, \bc_k \in C$ not all equal to $\ba^{(1)}$. This expression is nonzero since $C$ is average-free, implying $\beta(\bp, \bc)$ is nonzero.

In the degenerate case that $P_1 = P_{1,2}$, we have $t(P_1) = t(P_2)$ and $P_1 = P_1 \circ P_2[v,v]$. Otherwise, $P_{1,2}$ must at some point diverge from $P_1$ and thus have $\beta(\bp, \bc) \neq 0$ at that divergent step and each step afterwards. Since the sum of $R_{P_1, P_2}$ along the path is zero, this means that $\alpha(\bp, \bc)$ must be zero for the remainder of the path after diverging from $P_1$. But, since $\alpha$ is zero if and only if a step is in the direction of $P_2$ and $P_{1,2}$ ends at $t(P_2)$, this means that $P_{1,2}$ follows $P_2$ from the point at which it diverges from $P_1$. That is, $P_1$ and $P_2$ must intersect at the node at which the next step yields $\beta(\bp, \bc) > 0$. This shows $P_{1,2} = P_1[s(P_1), v] \circ P_2[v, t(P_2)]$ for some $v$, as required.
\end{proof}

\subsection{Finishing the proof of \Cref{thm:st-shortcut-lb}}

We have introduced all the new technical material we need to obtain polynomial lower bounds for Steiner shortcuts. Now we just need to put everything together, using standard arguments from the shortcut set lower bound literature. We begin by finalizing the construction details of our lower bound graph $G$ and critical paths $\mathcal{P}$. 

\paragraph{Specifying the parameters of our construction.} We choose our average-free set $C$ to be
$$C=\{(a,a^2):1\leq a\leq L^2\}\subseteq\ZZ^2.$$
It is straightforward to verify that $C$ is average-free by the strict convexity of $f(x) = x^2$.

Let $L\geq 3$ and let $k:=\dim(\bq)$, which is an absolute
constant since $d=2$ and $\Delta=4$.\footnote{In fact, $\bq$ has two entries for each monomial of degree at most four in the entries of $\bp$ and $\bc$. Then, since $\bp$ has $2d + 3 = 7$ entries and $\bc$ has $d = 2$ entries, there are $2{9 + 4 \choose 4} = 1430$ entries of $\bq$. That is, $k = 1430$.}
Set
\[
  R:=L^5,\qquad T:=L(2R)^4=16L^{21}.
\]
Define the source vertices $S$ and the full vertex set $V$ by
\begin{align*}
  S&:=\{1\}\times\{0,\ldots,R\}^{6}
               \times\{0,\ldots,T\}^{k},\\
  V&:=S\;\cup\;
       \bigl(\{2,\ldots,L\}\times\{0,\ldots,2R\}^{6}
                            \times\{0,\ldots,2T\}^{k}\bigr).
\end{align*}

The first coordinate of each vertex in $V$ corresponds to the current layer in $[1, L]$. The next six coordinates correspond to the two entries of $\br$ and the four entries of $\bx$. The final $k$ coordinates are used by $\bq$. 

Include each edge prescribed by our edge function~\eqref{eq:update} whenever both
endpoints belong to $V$, and let $\calP$ consist of the critical
paths starting in $S$, one for each pair $(\ba,\bb)\in C^2$.

Each non-layer position coordinate increases by at most $L^4$
per step, so a path starting in $S$ keeps these coordinates at
most $R+L\cdot L^4=2R$. All entries of $\bp$ and $\bc$ are
therefore bounded by $2R$. Every coordinate of $\upd_{\bq}$
is either zero or a monomial of degree at most four in these
entries, so it lies between zero and $(2R)^4$.
Thus each coordinate of $\bq$ is at most
$T+L(2R)^4=2T$. Then every critical path we defined
is contained in $G$.

We observe that $|S|=(R+1)^6(T+1)^k=\Theta(L^{30+21k})$.
Each layer of $G$ has $\Theta(|S|)$ vertices, and vertices in $G$ have on average $|C| = \Theta(L^2)$ outgoing edges. Since  every source and pair of directions define a distinct critical path, we obtain
\[
  n=\Theta(L |S|),\qquad
  m=\Theta(L |S| |C|),\qquad
  |\calP|= |S||C|^2=\Theta(mL).
\]
In particular, $n,m=O(L^K)$ for the fixed absolute constant
$K:=33+21k$.

\paragraph{Verifying Properties \ref{prop:unique} and \ref{prop:small-overlap}.}
Next we verify that our graph $G$ and critical paths $\mathcal{P}$ satisfy properties \ref{prop:unique} and \ref{prop:small-overlap}. These proofs do not depend on our choice of average-free set $C$.

\begin{claim} \label{clm:p1}
    The critical paths $\mathcal{P}$ satisfy property \ref{prop:unique}.
\end{claim}
\begin{proof}
Fix a critical path $P\in\calP$ with critical directions $(\ba,\bb)$, and let
$Q$ be any path between its endpoints in $G$. If
$\bc_1,\ldots,\bc_k\in C$ are the odd-step directions of path $Q$,
then our update rule~\eqref{eq:update} gives
\[
  \sum_{i=1}^k\bc_i
  =t(P)[\bx^{\odd}]-s(P)[\bx^{\odd}]
  =k \cdot \ba.
\]
Then the average-freeness of $C$ forces $\bc_i=\ba$ for every $i$. 
The same calculation for the even steps forces their 
directions to equal $\bb$. Since $P$ and $Q$ have the same starting point and direction vectors, we conclude $Q=P$.
\end{proof}

\begin{claim} \label{clm:p2}
    The critical paths $\mathcal{P}$ satisfy property \ref{prop:small-overlap} with $\rho = 1$. 
\end{claim}
\begin{proof}
Suppose towards contradiction that two distinct critical paths  $P, P' \in \mathcal{P}$ share two or more edges. Since critical paths are unique by~\Cref{clm:p1}, any two critical paths sharing at least two edges must share two \emph{consecutive} edges. Then these two paths have the same critical directions and pass through the same vertex. But paths in $\mathcal{P}$ are uniquely determined by their direction vectors and any individual vertex they pass through, so we conclude $P = P'$, a contradiction.
\end{proof}

We have shown that the critical paths $\mathcal{P}$ satisfy \ref{prop:unique} and \ref{prop:compatible-pairs}. Thus by  \Cref{thm:steiner-shortcut-template},  any Steiner shortcutting of our lower bound graph $G$ can be converted into a (non-Steiner) shortcutting with the same quality. Now we establish~\Cref{thm:st-shortcut-lb} by proving that $G$ does not admit a small low-diameter shortcut set using \ref{prop:small-overlap}. This argument is standard from prior work on shortcut sets due to~\cite{BH23}.
\PolynomialSteinerShortcutLowerBound*
\begin{proof}
Suppose that our lower bound graph $G$ admits a Steiner shortcutting $G_+$ with
$|E_+|\leq |\calP|/2$ edges and diameter $\diam(G_+)=D$.
By~\Cref{thm:steiner-shortcut-template}, there is a (non-Steiner) shortcutting $G'=(V,E\cup E')$ with $|E'|\leq |E_+|$ and
$d_{G'}(s(P),t(P))\leq D$ for every $P\in\calP$.

We will now prove that $D$ must be large. We will do this by analyzing a potential function $\phi$ that measures our progress towards reducing the diameter of $G$. Let $TC(G)$ be the transitive closure of $G$. For any shortcut set $H \subseteq TC(G)$, define the potential
\[
  \phi(H):=\sum_{P\in\calP}d_{G \cup H}(s(P),t(P)).
\]
By the pigeonhole principle, 
\[
D \geq \max_{P\in\calP} d_{G'}(s(P),t(P))
  \geq \frac{\phi(E')}{|\calP|}.
\]
Thus to lower bound the diameter of $G'$, it is sufficient to lower bound $\phi(E')$. We develop a few basic observations needed to prove this. First, note that $\phi(\emptyset)=(L-1)|\calP|$ since every path $P \in \calP$ has length $|P|=L-1$ in $G$.

Recall that $\rho$ is the path overlap parameter in property \ref{prop:small-overlap}.
Let $H_0 \subseteq TC(G)$ be the set of all edges $(u, v)$ in the transitive closure of $G$ with $\lay(v) - \lay(u) \le \rho+1$.  Then $H_0$ contains all edges in the transitive closure between vertex pairs that are at most $\rho+1$ layers apart. 
Since graph $G$ is layered and the paths in $\mathcal{P}$ pass through every layer of $G$, it follows that $$\phi(H_0) \ge \frac{L-1}{\rho+1} \cdot |\mathcal{P}|.$$

Now, for any edge $(u, v) \in TC(G)$, we want to bound how much the potential $\phi$ can decrease after adding edge $(u, v)$ to our shortcut set. We claim that, for any shortcut set $H \subseteq TC(G)$ and any edge $(u, v) \in TC(G)$,
\begin{equation}
    \phi(H \cup H_0)
    - \phi(H \cup H_0 \cup \{(u,v)\})
    \leq \left\lceil \frac{L-1}{\rho + 1} \right\rceil.
    \label{eq:potential-decrease}
\end{equation}
If $(u, v) \in H_0$, then this inequality is immediate. Otherwise, if $(u, v) \not \in H_0$, then by properties  \ref{prop:unique} and \ref{prop:small-overlap}, at most one path in $\mathcal{P}$ can contain both vertex $u$ and vertex $v$. Then shortcut edge $(u, v)$ can decrease the term $d_{G \cup H \cup H_0}(s(P), t(P))$ for at most one $P \in \mathcal{P}$. This path $P$ has length at most $\left\lceil(L-1)/(\rho+1) \right\rceil$ in  $G \cup H_0$, so \cref{eq:potential-decrease} follows.

We can now lower bound the diameter of the shortcutting $G'$. Let $E' = \{e_1, \dots, e_r\}$, and let $H_i := H_0\cup\{e_1,\ldots,e_i\}$ for $i \in [r]$. Then 
\begin{align*}
    \phi(E') & \ge \phi(E' \cup H_0) \\
    & = \phi(H_0) + \sum_{i \in [r]} \left( \phi(H_i) - \phi(H_{i-1})  \right) \\
    & \ge \frac{L-1}{\rho + 1} \cdot |\mathcal{P}| - \sum_{i \in [r]} \left\lceil\frac{L-1}{\rho + 1} \right\rceil\\
    & \ge \frac{L-1}{4(\rho + 1)} \cdot |\mathcal{P}|,
\end{align*}
where the final inequality follows from the fact that $r = |E'| \le |E_+| \le |\mathcal{P}|/2$. We conclude that $D \ge \phi(E') / |\mathcal{P}| \ge \Omega(L)$, using that our particular construction has $\rho = 1$. This completes the proof of \Cref{thm:st-shortcut-lb} by taking $\delta = 1/K$, where $K=O(1)$ is the constant we derived from our construction parameters.\footnote{Propagating through the value of $k = 1430$ gives $K = 30063$, resulting in $\delta = 1/30063$.}
\end{proof}

\section{Optimized Steiner Shortcut Lower Bounds} \label{sec:optimized}

In \Cref{sec:non-optimized}, we disproved the Steiner shortcut conjecture, \Cref{conjecture:steiner}. However, our lower bound only ruled out almost-linear size Steiner shortcuttings with diameter at most $n^{\delta}$ for $\delta =1/30063$. In this section, we improve this bound to $\delta = 1/56$, proving \Cref{thm:intro-shortcut-lb}.

\SteinerShortcutLowerBound*

% \begin{restatable}{theorem}{OptimizedSteinerShortcutLowerBound}
% \label{thm:optimized-shortcut-lb}
% There exists an infinite family of layered DAGs $G=(V,E)$ on $n$
% nodes and $m$ edges such that, for every $\varepsilon\ge0$, any
% Steiner shortcutting $G_+=(V\cup V_+,E\cup E_+)$ with
% $|E_+|=O(m^{1+\varepsilon})$ edges has $
%  \diam(G_+) \ge n^{1/56-O(\varepsilon)}.$
% \end{restatable}

Our optimized construction uses the same Steiner shortcut lower bound template that we saw in \Cref{sec:non-optimized}. In particular, we will guarantee that our lower bound graph $G$ and critical paths $\mathcal{P}$ satisfy properties \ref{prop:unique}, \ref{prop:compatible-pairs}, and \ref{prop:small-overlap}. As before, the main challenge will be to ensure property \ref{prop:compatible-pairs}, that pairs of compatible critical paths have a common vertex.

To obtain an improved lower bound, we must decrease the number of coordinates in the augmentation vector   $\bq$.  We achieve this by developing new optimized test functions that together prove the same splicing property \ref{prop:splicing-prop}  as in \Cref{sec:non-optimized}, but only use monomials of degree two (whereas \Cref{sec:non-optimized} used degree-four monomials). As a consequence, our new augmentation vector $\bq$ will use only ten coordinates, compared to the 1430 coordinates of $\bq$ in \Cref{sec:non-optimized}.

\subsection{Optimized test functions for splicing}
\label{subsec:optimized-splicing}

We begin by recalling the setup of \Cref{sec:non-optimized}. Each vertex $v \in V$ is a vector $\bv = (\bp, \bq)$, where the position vector $\bp = (\ell, \br, \bx)$ consists of the layer $\ell$, the counters $\br = (r_{\odd}, r_{\even})$, and $\bx = (\bx^{\odd}, \bx^{\even})$. We use the same update function
$ \upd(\bp, \bc)$ as in equation~\eqref{eq:update}.

Recall that for a compatible pair of paths $P_1, P_2 \in \calP$, a cross-path $P_{1,2}$ goes from $s(P_1)$ to $t(P_2)$, and a cross-path $P_{2,1}$ goes from $s(P_2)$ to $t(P_1)$. In \Cref{sec:non-optimized}, we transferred properties from paths $P_1$ and $P_2$ to their cross-paths using test functions $F_{P_1, P_2}$ with   properties~\ref{prop:nonneg}, \ref{prop:vanishing}, and~\ref{prop:linear}.

In this section, our test functions will only satisfy the following slightly weaker versions of properties~\ref{prop:nonneg} and~\ref{prop:linear}:
\begin{enumerate}[label = (T\arabic*$'$)]
 \item $F_{P_1,P_2}(\bp,\bc) \ge 0$ for all steps $(\bp,\bc)$ of the cross-paths $P_{1,2}$ and $P_{2,1}$. \label{prop:nonneg-cross}
 \setcounter{enumi}{2}
 \item $F_{P_1,P_2}(\bp,\bc)$ is a linear function of $\upd(\bp,\bc)$ plus a function of the layer $\bp[\ell]$; that is, there exist a coefficient vector $\bm{\lambda}_{P_1, P_2}$ and a function $g$ such that
 $$
 F_{P_1,P_2}(\bp,\bc) = \upd(\bp, \bc)^{\top} \bm{\lambda}_{P_1, P_2} + g(\bp[\ell]).
 $$
 \label{prop:linear-relaxed}
\end{enumerate}
Property~\ref{prop:linear-relaxed} differs from~\ref{prop:linear} in two ways. First, the linear function may use the whole update vector $\upd(\bp,\bc)$, which also contains the change of the position $\bp$, and not only its augmentation part $\upd_{\bq}(\bp,\bc)$. Second, we may add a function of the layer $\bp[\ell]$.

We use the extra freedom these weaker properties give us in order to develop test functions that add fewer coordinates to $\bq$. The following claim shows that these weaker properties are still sufficient for the argument of \Cref{sec:non-optimized}.
\begin{claim} \label{cla:relaxed}
Suppose that a function $F_{P_1,P_2}$ satisfies~\ref{prop:nonneg-cross}, \ref{prop:vanishing}, and~\ref{prop:linear-relaxed}. Then $F_{P_1,P_2}$ vanishes at every step of both cross-paths $P_{1,2}$ and $P_{2,1}$.
\end{claim}
\begin{proof}
Let $\blam_{P_1,P_2}$ and $g$ be as in~\ref{prop:linear-relaxed}. By~\eqref{eq:update}, each step from a vertex $\bv$ in direction $\bc$ adds the whole vector $\upd(\bv[\bp], \bc)$ to $\bv$. As in~\eqref{eq:use-of-linearity}, the sum of these vectors along a path $P$ is $t(P) - s(P)$. Moreover, a path from the first layer to the last layer takes exactly one step from each layer $\ell \in [L-1]$. Hence, for every such path $P$,
\[
 \sum_{(\bp,\bc)\text{ along }P} F_{P_1,P_2}(\bp,\bc) = \bigl(t(P) - s(P)\bigr)^{\top} \blam_{P_1,P_2} + \sum_{\ell=1}^{L-1} g(\ell).
\]
The cross-paths $P_{1,2}$ and $P_{2,1}$ together have the same start and end vertices as $P_1$ and $P_2$ together, and the extra $\sum_{\ell=1}^{L-1} g(\ell)$ term is the same for all four path sums since the graph is a layered DAG and all paths start and end on the same layers. Hence, the sum of $F_{P_1,P_2}$ along $P_1$ and $P_2$ equals its sum along $P_{1,2}$ and $P_{2,1}$.  By~\ref{prop:vanishing}, the sum along $P_1$ and $P_2$ is zero. By~\ref{prop:nonneg-cross}, every summand along $P_{1,2}$ and $P_{2,1}$ is nonnegative, so every summand is zero.
\end{proof}

The key property we want to give cross-paths is the splicing property~\ref{prop:splicing-prop},  which states that $P_{1,2} = P_1[s(P_1), v] \circ P_2[v, t(P_2)]$ for some vertex $v$ shared by paths $P_1$ and $P_2$.
In this subsection, we design three new test functions that combine to guarantee the splicing property~\ref{prop:splicing-prop}.

\paragraph{Guaranteeing the splicing property~\ref{prop:splicing-prop}.} Fix a compatible pair of paths $P_1, P_2 \in \mathcal{P}$ with critical directions $(\ba^{(1)}, \bb^{(1)})$ and $(\ba^{(2)}, \bb^{(2)})$ respectively. Likewise, fix a pair of cross-paths $P_{1,2}$ and $P_{2, 1}$.  We will build up to property~\ref{prop:splicing-prop} gradually, through the following three properties:
\begin{enumerate}[label = (Q\arabic*)]
 \item  Every odd step of a cross-path has direction $\ba^{(1)}$ or $\ba^{(2)}$, and every even step has direction $\bb^{(1)}$ or $\bb^{(2)}$. \label{prop:Q-labels}
 \item For cross-path $P_{1, 2}$,  all steps in direction $\ba^{(1)}$ come before all steps in direction $\ba^{(2)}$. Likewise, all steps in direction   $\bb^{(1)}$ come before all steps in direction $\bb^{(2)}$. For cross-path $P_{2,1}$, these orders are reversed. \label{prop:Q-order}
 \item There is a vertex $v$ in cross-path $P_{1, 2}$ such that the odd and even steps of path $P_{1, 2}[s(P_1), v]$ have directions $\ba^{(1)}$ and $\bb^{(1)}$, respectively, and the odd and even steps of $P_{1, 2}[v, t(P_2)]$ have directions $\ba^{(2)}$ and $\bb^{(2)}$, respectively. \label{prop:Q-sync}
\end{enumerate}

It is straightforward to verify that these three properties imply the splicing property for cross-path $P_{1, 2}$.
% \hfnoteinline{
% Currently you describe the Q’s as splitting (P2’) into three parts, but
% I’m wondering if the right way to say this is that we build up to (P2’) gradually by defining three test functions. The first one ensures (Q1), the second (Q2) using (Q1), and the third (P2’) using both (Q1) and (Q2). (In any case, the claim below is implied by just (Q3))}

\begin{claim} \label{cla:prefix-splice}
Suppose a cross-path $P_{1, 2}$ satisfies properties~\ref{prop:Q-labels}, \ref{prop:Q-order}, and \ref{prop:Q-sync}. Then $P_{1, 2}$ satisfies the splicing property \ref{prop:splicing-prop}.
\end{claim}

We will design one test function for each of the properties~\ref{prop:Q-labels}, \ref{prop:Q-order}, and \ref{prop:Q-sync}. Unlike in \Cref{sec:non-optimized}, our test functions will explicitly depend on the construction details of our graph $G$ and critical paths $\calP$. Therefore, we introduce some of these details before proceeding.

\paragraph{Construction details for $G$ and $\calP$.}
As in \Cref{sec:non-optimized}, the set of direction vectors $C$ must be average-free, which gives properties~\ref{prop:unique} and~\ref{prop:small-overlap}. Our set $C$ will contain $q$ vectors $\bc_0, \bc_1, \dots, \bc_{q-1}$, where $q$ is a prime with $L \le q \le 2L$ (such a $q$ exists by, e.g., Bertrand's postulate). For each $t \in [0, q-1]$, we say that $t$ is the \emph{label} of vector $\bc_t \in C$. Let
$$
y_t := t^2 \mod q \qquad \text{ and } \qquad z_t := (t^2 - y_t)/q,
$$
so that $t^2 = qz_t + y_t$, and let
\begin{equation}
 \bc_t := (t,\ y_t,\ z_t,\ t y_t) \in \ZZ^4  \text{ for } t \in [0, q-1].
\end{equation}
The set $C = \{\bc_0, \bc_1, \dots, \bc_{q-1}\}$ is average-free. The linear map $(u_1,u_2,u_3,u_4) \mapsto (u_1, qu_3 + u_2)$ sends $\bc_t$ to the point $(t,t^2)$, so if $k\bc_{t_0} = \bc_{t_1} + \cdots + \bc_{t_k}$, then $k(t_0,t_0^2) = (t_1,t_1^2) + \cdots + (t_k,t_k^2)$, which implies $t_1 = \cdots = t_k = t_0$ because $f(t) = t^2$ is strictly convex.

We note that our set $C$ uses more coordinates than average-freeness requires. These extra coordinates are quantities that our test functions need to track along a path; placing them in the directions $C$ lets the position $\bp$ track them, which saves coordinates of $\bq$.

We also make two changes to the construction of \Cref{sec:non-optimized}. First, to ensure property~\ref{prop:Q-order}, the counters of every source vertex are zero. That is, every source vertex $s \in S$ satisfies $s[r_{\odd}] = s[r_{\even}] = 0$. (In \Cref{sec:non-optimized}, the counters of the source vertices ranged over $\{0,\ldots,L^5\}$.) Since each step increases one of the two counters by one, the counters of every vertex are then determined by its layer. Second, to ensure property~\ref{prop:Q-sync}, the number of layers $L$ is even.

\paragraph{Ensuring property \ref{prop:Q-labels}.}
For our compatible pair of paths $P_1, P_2 \in \mathcal{P}$, let $a_i, b_i \in [0, q-1]$ be the labels of the direction vectors $\ba^{(i)}, \bb^{(i)} \in C$, for $i \in [1, 2]$.
 Property \ref{prop:Q-labels} claims that each odd step of a cross-path has label $a_1$ or $a_2$, and each even step has label $b_1$ or $b_2$.

 In order to restrict the labels of the odd steps, we desire a nonnegative test function that vanishes exactly at the two labels $a_1$ and $a_2$. To define our test function, we will use the line through the points $(a_1, y_{a_1})$ and $(a_2, y_{a_2})$. Let $y = \mu_{\odd} t + \nu_{\odd}$ be this line if $a_1 \ne a_2$. Else if $a_1 = a_2$, we instead let $y = \mu_{\odd} t + \nu_{\odd}$ be the line $y = y_{a_1} + 2a_1(t - a_1)$, which passes through $(a_1, y_{a_1})$ with slope $2a_1$. We define $\mu_{\even}$ and $\nu_{\even}$ in the same way from the labels $b_1$ and $b_2$. Our test function for property \ref{prop:Q-labels} is
\begin{equation}
 Y_{P_1,P_2}(\bp, \bc_t) := \begin{cases}
 (y_t - \mu_{\odd} t - \nu_{\odd})^2 & \text{if } \bp[\ell] \equiv 1 \pmod 2, \\
 (y_t - \mu_{\even} t - \nu_{\even})^2 & \text{if } \bp[\ell] \equiv 0 \pmod 2.
 \end{cases} \label{eq:label-test}
\end{equation}
At an odd step, we can interpret $Y_{P_1,P_2}$ as the squared vertical distance from the point $(t, y_t)$ of the current label to the line through the points $(a_1, y_{a_1})$ and $(a_2, y_{a_2})$. Thus, it is zero exactly when the point of the current label lies on this line. This gives property~\ref{prop:Q-labels}, as stated in the following claim.

\begin{restatable}{claim}{LabelsClaim} \label{cla:labels}
Every odd step of a cross-path $P_{1,2}$ or $P_{2,1}$ has label $a_1$ or $a_2$, and every even step has label $b_1$ or $b_2$.
\end{restatable}
Before we prove \Cref{cla:labels}, we verify that $Y_{P_1,P_2}$ has properties~\ref{prop:nonneg-cross}, \ref{prop:vanishing}, and~\ref{prop:linear-relaxed}.  The function $Y_{P_1,P_2}$ is clearly nonnegative, so it satisfies~\ref{prop:nonneg}, and hence~\ref{prop:nonneg-cross}. Additionally,  $Y_{P_1,P_2}$ vanishes at every step of $P_1$ and $P_2$, because the points $(a_1, y_{a_1})$ and $(a_2, y_{a_2})$ lie on the line $y = \mu_{\odd} t + \nu_{\odd}$, and the points $(b_1, y_{b_1})$ and $(b_2, y_{b_2})$ lie on the line $y = \mu_{\even} t + \nu_{\even}$. This gives property~\ref{prop:vanishing}.

% \begin{claim} \label{cla:relaxed}
% Suppose that a function $F_{P_1,P_2}$ satisfies~\ref{prop:nonneg}, \ref{prop:vanishing}, and~\ref{prop:linear-relaxed}. Then $F_{P_1,P_2}$ vanishes at every step of both cross paths $P_{1,2}$ and $P_{2,1}$.
% \end{claim}
% \begin{proof}
% Let $\blam_{P_1,P_2}$ and $g$ be as in~\ref{prop:linear-relaxed}. By~\eqref{eq:update}, each step from a vertex $\bv$ in direction $\bc$ adds the whole vector $\upd(\bv[\bp], \bc)$ to $\bv$. As in~\eqref{eq:use-of-linearity}, the sum of these vectors along a path $P$ is $t(P) - s(P)$. Moreover, a path from the first layer to the last layer takes exactly one step from each layer $\ell \in [L-1]$. Hence, for every such path $P$,
% \[
%  \sum_{(\bp,\bc)\text{ along }P} F_{P_1,P_2}(\bp,\bc) = \bigl(t(P) - s(P)\bigr)^{\top} \blam_{P_1,P_2} + \sum_{\ell=1}^{L-1} g(\ell).
% \]
% The cross paths $P_{1,2}$ and $P_{2,1}$ together have the same start and end vertices as $P_1$ and $P_2$ together, and the extra $\sum_{\ell=1}^{L-1} g(\ell)$ term is the same for all four path sums since the graph is a layered DAG and all paths start and end on the same layers. Hence, the sum of $F_{P_1,P_2}$ along $P_1$ and $P_2$ equals its sum along $P_{1,2}$ and $P_{2,1}$.  By~\ref{prop:vanishing}, the sum along $P_1$ and $P_2$ is zero. By~\ref{prop:nonneg}, every summand along $P_{1,2}$ and $P_{2,1}$ is nonnegative, so every summand is zero.
% \end{proof}
% \ghnote{This proof for Claim 3.2 may be overkill}

To guarantee that $Y_{P_1,P_2}$ satisfies~\ref{prop:linear-relaxed}, we add one coordinate to $\bq$ that records $y_t^2$ at every step. This simple addition to $\bq$ is enough to ensure that~\ref{prop:linear-relaxed} holds, as we now prove.
\begin{claim} \label{cla:Y-linear}
The test function $Y_{P_1,P_2}$ satisfies~\ref{prop:linear-relaxed}.
\end{claim}
\begin{proof}
Consider a step $(\bp, \bc_t)$, and let $\sigma \in \{\odd, \even\}$ be the parity of its layer $\bp[\ell]$. Expanding $ Y_{P_1,P_2}(\bp, \bc_t)$ and using $t^2 = qz_t + y_t$, we get
\[
 Y_{P_1,P_2}(\bp, \bc_t) = y_t^2 + \bigl(2\mu_\sigma\nu_\sigma\, t + (\mu_\sigma^2 - 2\nu_\sigma)\, y_t + q\mu_\sigma^2\, z_t - 2\mu_\sigma\, t y_t\bigr) + \nu_\sigma^2.
\]
The term $y_t^2$ is recorded by the new coordinate of $\upd_{\bq}(\bp,\bc_t)$. The middle term is a linear combination of the entries of $\bc_t = (t, y_t, z_t, ty_t)$, which is the change of $\bx^{\sigma}$ at this step. The constant $\nu_\sigma^2$ is $\nu_\sigma^2$ times the change of $r_{\sigma}$, which is one at this step. We therefore let the coefficient vector $\blam_{P_1,P_2}$ have entry $1$ at the new coordinate of $\bq$ and, for both parities $\sigma \in \{\odd, \even\}$,
\[
 \blam_{P_1,P_2}[\bx^{\sigma}] := \bigl(2\mu_\sigma\nu_\sigma,\ \mu_\sigma^2 - 2\nu_\sigma,\ q\mu_\sigma^2,\ -2\mu_\sigma\bigr), \qquad \blam_{P_1,P_2}[r_{\sigma}] := \nu_\sigma^2,
\]
with all other entries zero. 
A step whose layer has parity $\sigma$ does not change $\bx^{\sigma'}$ or $r_{\sigma'}$ for the other parity $\sigma'$, so $\upd(\bp,\bc_t)^{\top}\blam_{P_1,P_2}$ is exactly the expansion above. Thus the same coefficient vector works for both kinds of steps, and $Y_{P_1,P_2}$ needs no coordinate of $\bq$ other than the one that records $y_t^2$. This gives~\ref{prop:linear-relaxed}, with $g = 0$.
\end{proof}

By \Cref{cla:relaxed,cla:Y-linear}, $Y_{P_1,P_2}$ vanishes at every step of both cross-paths. We can now prove that $Y_{P_1,P_2}$ ensures property~\ref{prop:Q-labels}:

\LabelsClaim*
\begin{proof}[Proof]
Since $Y_{P_1,P_2}$ vanishes at each step of both cross-paths, every odd step of a cross-path has a label $t$ such that $(t, y_t)$ lies on the line $y = \mu_{\odd} t + \nu_{\odd}$. We show that $a_1$ and $a_2$ are the only labels with this property. Suppose first that $a_1 \ne a_2$. Then this line passes through the points $(a_1, y_{a_1})$ and $(a_2, y_{a_2})$, so this line can be written in its two-point form as
\[
 y - y_{a_1} = \frac{y_{a_2} - y_{a_1}}{a_2 - a_1}\,(t - a_1).
\]
If $(t, y_t)$ lies on this line, then multiplying both sides by $a_2 - a_1$ to clear the denominator gives
\[
 (y_t - y_{a_1})(a_2 - a_1) = (y_{a_2} - y_{a_1})(t - a_1).
\]
Then taking this equation modulo $q$ and using the fact that $y_s \equiv s^2 \pmod q$ for every label $s$, we get $(t^2 - a_1^2)(a_2 - a_1) \equiv (a_2^2 - a_1^2)(t - a_1)$. Rearranging and factoring, we obtain $$(t-a_1)(t-a_2)(a_2-a_1) \equiv 0 \pmod q.$$ Since $q$ is prime and does not divide $a_2 - a_1$, we get $t \equiv a_1$ or $t \equiv a_2 \pmod q$, and hence $t \in \{a_1, a_2\}$, as all labels lie in $[0, q-1]$. If $a_1 = a_2$, then a point $(t, y_t)$ on the line $y = y_{a_1} + 2a_1(t - a_1)$ satisfies $t^2 \equiv a_1^2 + 2a_1(t - a_1)$, that is, $(t - a_1)^2 \equiv 0 \pmod q$, so again $t = a_1$. The even steps are handled in the same way, with $b_1$, $b_2$, $\mu_{\even}$, and $\nu_{\even}$.
\end{proof}

\paragraph{Ensuring property~\ref{prop:Q-order}.}
Recall that property~\ref{prop:Q-order} requires that on cross-path $P_{1,2}$, all steps in direction $\ba^{(1)}$ come before all steps in direction $\ba^{(2)}$, and all steps in direction $\bb^{(1)}$ come before all steps in direction $\bb^{(2)}$, while on $P_{2,1}$, these orders are reversed. We consider the odd steps; the even steps are handled in the same way. If $a_1 = a_2$, then, by \Cref{cla:labels}, every odd step of a cross-path has label $a_1 = a_2$, so the order of the odd steps does not matter. We therefore assume that $a_1 \ne a_2$.

Our test function for property~\ref{prop:Q-order} on the odd steps is the product of two factors $\psi_1$ and $\psi_2$. At an odd step of cross-path $P_{1,2}$, the factor $\psi_1$ counts the earlier odd steps in direction $\ba^{(2)}$, and $\psi_2$ counts the odd steps in direction $\ba^{(1)}$ that are still to come, including the current step. Hence $\psi_1$ is zero up to the first odd step in direction $\ba^{(2)}$, and $\psi_2$ is zero after the last odd step in direction $\ba^{(1)}$. Their product is zero at every odd step exactly when no odd step in direction $\ba^{(1)}$ comes after an odd step in direction $\ba^{(2)}$. This will give  property~\ref{prop:Q-order}, as we prove in \Cref{cla:order}.

To define $\psi_1$ and $\psi_2$, recall that the counters of every source vertex are zero. For a position $\bp$, write $x(\bp) := \bp[\bx^{\odd}]_1$ for the first entry of $\bp[\bx^{\odd}]$, and $r(\bp) := \bp[r_{\odd}]$. Since the source counters are zero, $r(\bp)$ is the number of odd steps before position $\bp$, so it depends only on the layer $\bp[\ell]$. Similarly, since the first entry of each direction $\bc_t$ is its label $t$, on a path from a source vertex $s$, $x(\bp)$ is $x(s)$ plus the sum of the labels of the odd steps before position $\bp$.  Let $x_i := x(s(P_i))$ be the starting value of $x$ on $P_i$. Since every odd step of $P_i$ has label $a_i$, the vertex of $P_i$ in layer $\bp[\ell]$ has $x$-value $x_i + a_i\, r(\bp)$. For $i \in \{1,2\}$, the factor $\psi_i$ compares $x(\bp)$ with this value:
\[
 \psi_i(\bp) := \frac{x(\bp) - x_i - a_i\, r(\bp)}{a_2 - a_1}.
\]
Our test function for property~\ref{prop:Q-order} on the odd steps is
\begin{equation}
 R^{\odd}_{P_1,P_2}(\bp,\bc) := \begin{cases}
 \psi_1(\bp)\,\psi_2(\bp) & \text{if } \bp[\ell] \equiv 1 \pmod 2,\\
 0 & \text{if } \bp[\ell] \equiv 0 \pmod 2.
 \end{cases} \label{eq:R-odd}
\end{equation}
The test function $R^{\even}_{P_1,P_2}$ for the even steps is defined in the same way from $b_1$, $b_2$, $\bx^{\even}$, and $r_{\even}$.

We now check that $R^{\odd}_{P_1,P_2}$ and $R^{\even}_{P_1,P_2}$ satisfy properties~\ref{prop:nonneg-cross}, \ref{prop:vanishing}, and~\ref{prop:linear-relaxed}. We do this for $R^{\odd}_{P_1,P_2}$; the argument for $R^{\even}_{P_1,P_2}$ is the same. An odd step with label $t$ adds $t$ to $x(\bp)$ and one to $r(\bp)$, so it changes $\psi_i$ by $(t - a_i)/(a_2 - a_1)$, and even steps do not change $\psi_i$. Along $P_i$, every odd step has label $a_i$, so $\psi_i$ keeps its value $0$ at $s(P_i)$. Hence $R^{\odd}_{P_1,P_2}$ vanishes at every step of $P_1$ and $P_2$, which gives~\ref{prop:vanishing}.

By \Cref{cla:labels}, along a cross-path, $\psi_1$ increases by one at each odd step with label $a_2$, $\psi_2$ decreases by one at each odd step with label $a_1$, and no other step changes them. On $P_{1,2}$, $\psi_1$ is $0$ at the start $s(P_1)$ and $\psi_2$ is $0$ at the end $t(P_2)$, so both are nonnegative along $P_{1,2}$. In particular, along $P_{1,2}$, the factors $\psi_1$ and $\psi_2$ count steps as described above. On $P_{2,1}$, $\psi_2$ is $0$ at the start $s(P_2)$ and $\psi_1$ is $0$ at the end $t(P_1)$, so both are nonpositive along $P_{2,1}$. Hence $R^{\odd}_{P_1,P_2}$ is nonnegative at every step of a cross-path, which gives~\ref{prop:nonneg-cross}. (It can be negative at steps that do not lie on a cross-path, which is why we need the weaker property~\ref{prop:nonneg-cross} in place of~\ref{prop:nonneg}.)

To guarantee that $R^{\odd}_{P_1,P_2}$ and $R^{\even}_{P_1,P_2}$ satisfy the relaxed linearity property~\ref{prop:linear-relaxed}, we add six coordinates to $\bq$. Three of them record $x(\bp)$, $r(\bp)\,x(\bp)$, and $x(\bp)^2$ at each odd step $(\bp,\bc)$, and zero at even steps. The other three record the analogous quantities $\bp[\bx^{\even}]_1$, $\bp[r_{\even}]\,\bp[\bx^{\even}]_1$, and $\bp[\bx^{\even}]_1^2$ at each even step, and zero at odd steps. These coordinates are enough to ensure that~\ref{prop:linear-relaxed} holds, as we now prove.
\begin{claim} \label{cla:R-linear}
The test functions $R^{\odd}_{P_1,P_2}$ and $R^{\even}_{P_1,P_2}$ satisfy~\ref{prop:linear-relaxed}.
\end{claim}
\begin{proof}
We prove this for $R^{\odd}_{P_1,P_2}$; the proof for $R^{\even}_{P_1,P_2}$ is the same. Consider an odd step $(\bp,\bc)$. Multiplying out $\psi_1(\bp)\,\psi_2(\bp)$, we get
\[
 R^{\odd}_{P_1,P_2}(\bp,\bc) = \frac{x(\bp)^2 - (a_1 + a_2)\, r(\bp)\,x(\bp) - (x_1 + x_2)\, x(\bp) + \bigl(x_1 + a_1 r(\bp)\bigr)\bigl(x_2 + a_2 r(\bp)\bigr)}{(a_2 - a_1)^2}.
\]
The first three terms of the numerator are constant multiples of $x(\bp)^2$, $r(\bp)\,x(\bp)$, and $x(\bp)$, which are recorded by three new coordinates of $\upd_{\bq}(\bp,\bc)$. The last term depends only on $r(\bp)$, and hence only on the layer $\bp[\ell]$. Hence, $R^{\odd}_{P_1,P_2}(\bp,\bc)$ can be written as $\upd(\bp,\bc)^{\top}\blam + g(\bp[\ell])$. This gives~\ref{prop:linear-relaxed}.
% , 
% where $\blam$ has the entries $1/(a_2-a_1)^2$, $-(a_1+a_2)/(a_2-a_1)^2$, and $-(x_1+x_2)/(a_2-a_1)^2$ at these three coordinates and zero elsewhere, and $g(\ell)$ is the last term divided by $(a_2-a_1)^2$, with $r(\bp)$ replaced by its value in layer $\ell$. At an even step, $R^{\odd}_{P_1,P_2}$ is zero, and so are these three coordinates, so the same identity holds if we let $g(\ell) := 0$ for even $\ell$. This gives~\ref{prop:linear-relaxed}.
\end{proof}

By \Cref{cla:relaxed,cla:R-linear}, $R^{\odd}_{P_1,P_2}$ and $R^{\even}_{P_1,P_2}$ vanish at every step of both cross-paths. We can now prove that they ensure property~\ref{prop:Q-order}:

\begin{claim} \label{cla:order}
Suppose that $a_1 \ne a_2$. On cross-path $P_{1,2}$, all odd steps in direction $\ba^{(1)}$ come before all odd steps in direction $\ba^{(2)}$, and on $P_{2,1}$, this order is reversed. Likewise, if $b_1 \ne b_2$, then on $P_{1,2}$, all even steps in direction $\bb^{(1)}$ come before all even steps in direction $\bb^{(2)}$, and on $P_{2,1}$, this order is reversed.
\end{claim}
\begin{proof}
We prove the statement for the odd steps; the proof for the even steps is the same. A step is in direction $\ba^{(i)}$ exactly when it has label $a_i$. Since $R^{\odd}_{P_1,P_2}$ vanishes at each step of both cross-paths, we have $\psi_1(\bp) = 0$ or $\psi_2(\bp) = 0$ at each odd step $(\bp,\bc)$ of a cross-path. Suppose that some odd step of $P_{1,2}$ with label $a_2$ comes before an odd step with label $a_1$. At the later step, we have $\psi_1(\bp) \ge 1$, which counts the earlier odd step with label $a_2$, and $\psi_2(\bp) \ge 1$, which counts the current step with label $a_1$. This contradicts the vanishing of $R^{\odd}_{P_1,P_2}$, so on $P_{1,2}$, all odd steps with label $a_1$ come before all odd steps with label $a_2$. Similarly, if an odd step of $P_{2,1}$ with label $a_1$ came before an odd step with label $a_2$, then at the later step, we would have $\psi_2(\bp) \le -1$ and $\psi_1(\bp) \le -1$. Hence on $P_{2,1}$, all odd steps with label $a_2$ come before all odd steps with label $a_1$.
\end{proof}

\paragraph{Ensuring property~\ref{prop:Q-sync}.}
Recall that property~\ref{prop:Q-sync} requires that cross-path $P_{1,2}$ first takes the directions $\ba^{(1)}$ and $\bb^{(1)}$ of $P_1$ at its odd and even steps, respectively, and then the directions $\ba^{(2)}$ and $\bb^{(2)}$ of $P_2$. By \Cref{cla:order}, within each parity, $P_{1,2}$ first takes the direction of $P_1$ and then the direction of $P_2$. But the odd steps and the even steps could switch at different points. We now design a test function that rules this out.

We first suppose that $a_1 \ne a_2$ and $b_1 \ne b_2$; the cases with equal labels are handled at the end. For $\sigma\in\{\odd,\even\}$, write $x^{\sigma}(\bp):=\bp[\bx^{\sigma}]_1$, $r^{\sigma}(\bp):=\bp[r_{\sigma}]$, and $x_i^{\sigma}:=s(P_i)[\bx^{\sigma}]_1$. Define
\[
 \tau_{\odd} := \frac{x_1^{\odd}-x_2^{\odd}}{a_2-a_1},
 \qquad
 \tau_{\even} := \frac{x_1^{\even}-x_2^{\even}}{b_2-b_1}.
\]
These quantities count the initial steps in each parity before the switch. (Think $\tau$ for \emph{threshold}.) Recall that $L$ is even, so every path from the first layer to the last layer has $L/2$ odd steps and $L/2-1$ even steps. If $P_{1,2}$ takes $\tau$ odd steps with label $a_1$, comparing its endpoints gives
\[
 x_1^{\odd} + \tau a_1 + (L/2-\tau)a_2 = x_2^{\odd} + (L/2)a_2,
\]
so $\tau=\tau_{\odd}$. The same comparison for $P_{2,1}$ shows that it takes $\tau_{\odd}$ odd steps with label $a_2$. Similarly, $P_{1,2}$ takes $\tau_{\even}$ even steps with label $b_1$, and $P_{2,1}$ takes $\tau_{\even}$ even steps with label $b_2$. By \Cref{cla:order}, these are the first steps of their respective parities.

Let $\psi_2^{\even}$ be the even analogue of the function $\psi_2$ used for property~\ref{prop:Q-order}; explicitly,
\[
 \psi_2^{\even}(\bp) := \frac{x^{\even}(\bp)-x_2^{\even}-b_2r^{\even}(\bp)}{b_2-b_1}.
\]
Our test function for property~\ref{prop:Q-sync} is
\begin{equation}
 S_{P_1,P_2}(\bp,\bc_t) := \begin{cases}
 (\tau_{\even}-\tau_{\odd})\,\dfrac{t-a_1}{a_2-a_1}\,\psi_2^{\even}(\bp)
 & \text{if } \bp[\ell] \equiv 1 \pmod 2,\\
 0 & \text{if } \bp[\ell] \equiv 0 \pmod 2.
 \end{cases} \label{eq:S-test}
\end{equation}
Intuitively, on $P_{1,2}$, the factor $(t-a_1)/(a_2-a_1)$ detects whether the current odd step has already switched to label $a_2$, while $\psi_2^{\even}$ counts the even steps with label $b_1$ still to come. Their product then detects an odd step that switches too early. On $P_{2,1}$, the same product detects an odd step that switches too late, with the opposite sign. The factor $\tau_{\even}-\tau_{\odd}$ makes the test nonnegative in both cases.

We now check the properties of this test function. Along $P_1$, every odd step has label $a_1$, so the factor $t-a_1$ is zero. Along $P_2$, the function $\psi_2^{\even}$ is zero. Hence $S_{P_1,P_2}$ vanishes at every step of $P_1$ and $P_2$, which gives~\ref{prop:vanishing}.

To check property~\ref{prop:nonneg-cross}, number the odd steps by $j=0,1,\ldots,L/2-1$. At odd step $j$ of $P_{1,2}$, the factor $(t-a_1)/(a_2-a_1)$ is one if $j\ge\tau_{\odd}$ and zero otherwise, and $\psi_2^{\even}$ is $\max\{\tau_{\even}-j,0\}$. Their product is nonzero (and positive) only if $\tau_{\odd}\le j<\tau_{\even}$, in which case $\tau_{\even}-\tau_{\odd}>0$. Thus $S_{P_1,P_2}$ is nonnegative along $P_{1,2}$.

At odd step $j$ of $P_{2,1}$, the first factor is one if $j<\tau_{\odd}$ and zero otherwise, and $\psi_2^{\even}$ is $-\max\{j-\tau_{\even},0\}$. Their product is nonzero (and negative) only if $\tau_{\even}<j<\tau_{\odd}$, in which case $\tau_{\even}-\tau_{\odd}<0$. Thus $S_{P_1,P_2}$ is also nonnegative along $P_{2,1}$, giving~\ref{prop:nonneg-cross}.

To guarantee that $S_{P_1,P_2}$ satisfies the relaxed linearity property~\ref{prop:linear-relaxed}, we add three coordinates to $\bq$ that record $t\,x^{\even}(\bp)$, $x^{\even}(\bp)$, and $r^{\odd}(\bp)t$ at each odd step $(\bp,\bc_t)$, and zero at even steps. At an odd step, we have $r^{\odd}(\bp)=r^{\even}(\bp)$, so
\begin{align*}
 S_{P_1,P_2}(\bp,\bc_t)
 &= \frac{\tau_{\even}-\tau_{\odd}}{(a_2-a_1)(b_2-b_1)}
 \Bigl(t x^{\even}(\bp)-a_1x^{\even}(\bp)-b_2r^{\odd}(\bp)t\\
 &\hspace{42mm}{}-x_2^{\even}t+a_1x_2^{\even}+a_1b_2r^{\odd}(\bp)\Bigr).
\end{align*}
The first three terms are linear in the three new coordinates of $\upd_{\bq}(\bp,\bc_t)$. The fourth term is $-x_2^{\even}$ times the change in $x^{\odd}(\bp)$, which is also a coordinate of $\upd(\bp,\bc_t)$. The last two terms depend only on the layer. At an even step, the three new coordinates and the change in $x^{\odd}(\bp)$ are all zero, so the same representation holds with the layer-dependent term set to zero. Thus $S_{P_1,P_2}(\bp,\bc)$ can be written as $\upd(\bp,\bc)^{\top}\blam + g(\bp[\ell])$, giving~\ref{prop:linear-relaxed}.

By \Cref{cla:relaxed}, using properties~\ref{prop:nonneg-cross}, \ref{prop:vanishing}, and~\ref{prop:linear-relaxed}, $S_{P_1,P_2}$ vanishes at every step of both cross-paths. We can now prove that this ensures property~\ref{prop:Q-sync}:
\begin{claim} \label{cla:sync}
The cross-path $P_{1,2}$ satisfies property~\ref{prop:Q-sync}.
\end{claim}
\begin{proof}
Suppose first that $a_1\ne a_2$ and $b_1\ne b_2$. If $\tau_{\odd}<\tau_{\even}$, then at odd step $j=\tau_{\odd}$ of $P_{1,2}$, the test function has value $(\tau_{\even}-\tau_{\odd})^2>0$. This step exists because $\tau_{\odd}<\tau_{\even}\le L/2$, and it contradicts the vanishing of $S_{P_1,P_2}$. If $\tau_{\odd}>\tau_{\even}+1$, then at odd step $j=\tau_{\even}+1$ of $P_{2,1}$, the test function has value $\tau_{\odd}-\tau_{\even}>0$, again a contradiction. This step exists because $\tau_{\even}+1<\tau_{\odd}\le L/2$. We conclude that
\begin{equation}
 \tau_{\odd}=\tau_{\even}
 \qquad\text{or}\qquad
 \tau_{\odd}=\tau_{\even}+1. \label{eq:sync-cond}
\end{equation}
In the first case, let $v$ be the vertex after the first $2\tau_{\odd}$ steps of $P_{1,2}$; in the second, let $v$ be the vertex after the first $2\tau_{\odd}-1$ steps. By \Cref{cla:order}, all steps before $v$ have the directions of $P_1$, and all steps after $v$ have the directions of $P_2$, as required.

It remains to handle the cases with equal labels. If $a_1=a_2$ and $b_1=b_2$, every step has the direction of both paths, so we can take $v=s(P_1)$. If $a_1=a_2$ but $b_1\ne b_2$, take $v$ immediately after the last even step with label $b_1$, or at $s(P_1)$ if there is no such step. By \Cref{cla:order}, this gives the required split. If $a_1\ne a_2$ but $b_1=b_2$, take $v$ immediately after the last odd step with label $a_1$, or at $s(P_1)$ if there is no such step.
\end{proof}

\paragraph{Putting the three properties together.}
We can now combine~\ref{prop:Q-labels}, \ref{prop:Q-order}, and~\ref{prop:Q-sync} to prove the splicing property for our new construction.
\begin{claim} \label{cla:splice}
Let $P_1, P_2$ be a compatible pair. Then every cross-path $P_{1,2}$ satisfies $P_{1,2} = P_1[s(P_1), v] \circ P_2[v, t(P_2)]$ for some vertex $v$ of both $P_1$ and $P_2$. That is, our construction has the splicing property~\ref{prop:splicing-prop}.
\end{claim}
\begin{proof}
By \Cref{cla:labels,cla:order,cla:sync}, $P_{1,2}$ has properties~\ref{prop:Q-labels}, \ref{prop:Q-order}, and~\ref{prop:Q-sync}, so \Cref{cla:prefix-splice} gives the claim.
\end{proof}

\subsection{Finishing the proof of \Cref{thm:intro-shortcut-lb}}
\label{subsec:finish}

We have shown how to ensure property~\ref{prop:compatible-pairs} with our test functions. Now we put everything together. As in \Cref{sec:non-optimized}, we first specify the parameters of our lower bound graph $G$ and its critical paths $\calP$, and then we verify properties~\ref{prop:unique}, \ref{prop:compatible-pairs}, and~\ref{prop:small-overlap}.

\paragraph{Specifying the parameters of our construction.}
Recall that $q$ is a prime with $L \le q \le 2L$, and that the directions $\bc_t = (t, y_t, z_t, ty_t)$ are indexed by the labels $t \in [0, q-1]$.

Collecting the coordinates that we added in \Cref{subsec:optimized-splicing}, the augmentation vector $\bq$ has exactly the following ten coordinates:
\begin{enumerate}
 \item one coordinate for the test function $Y_{P_1,P_2}$ of property~\ref{prop:Q-labels}, which records $y_t^2$ at each step;
 \item three coordinates for the test function $R^{\odd}_{P_1,P_2}$ of property~\ref{prop:Q-order}, which record $\bp[\bx^{\odd}]_1$, $\bp[r_{\odd}]\,\bp[\bx^{\odd}]_1$, and $\bp[\bx^{\odd}]_1^2$ at odd steps, and three coordinates for $R^{\even}_{P_1,P_2}$, which record $\bp[\bx^{\even}]_1$, $\bp[r_{\even}]\,\bp[\bx^{\even}]_1$, and $\bp[\bx^{\even}]_1^2$ at even steps;
 \item three coordinates for the test function $S_{P_1,P_2}$ of property~\ref{prop:Q-sync}, which record $t\cdot\bp[\bx^{\even}]_1$, $\bp[\bx^{\even}]_1$, and $\bp[r_{\odd}]\cdot t$ at odd steps.
\end{enumerate}

For the entries of $\bx^{\odd}$ and $\bx^{\even}$, set $X := Lq$ and $X' := Lq^2$. For the coordinates of $\bq$, set
\begin{gather*}
 B_1 := Lq^2, \qquad B_2 := 2L^2q, \qquad B_3 := 2L^3q, \qquad B_4 := 4L^3q^2,\\
 B_5 := 2L^2q^2, \qquad B_6 := 2L^2q, \qquad B_7 := L^2q.
\end{gather*}
Define the source vertices $S$ and the full vertex set $V$ by
\begin{align*}
 S &:= \{(1,0,0)\} \times \bigl([0,X]^3\times[0,X']\bigr)^2 \times [0,B_1] \times \bigl([0,B_2]\times[0,B_3]\times[0,B_4]\bigr)^2\\
 &\qquad \times [0,B_5] \times [0,B_6] \times [0,B_7],\\
 \vspace{3mm}
 \\
 V &:= S \;\cup\; \bigcup_{\ell=2}^{L} \Bigl(\bigl\{\bigl(\ell, \lfloor \ell/2 \rfloor, \lfloor (\ell-1)/2 \rfloor\bigr)\bigr\} \times \bigl([0,2X]^3\times[0,2X']\bigr)^2\\
 &\qquad \times [0,2B_1] \times \bigl([0,2B_2]\times[0,2B_3]\times[0,2B_4]\bigr)^2 \times [0,2B_5] \times [0,2B_6] \times [0,2B_7]\Bigr).
\end{align*}

The first three coordinates of each vertex in $V$ are its layer $\ell$ and its counters $r_{\odd}$ and $r_{\even}$. Since the source counters are zero, the counters at layer $\ell$ are the numbers $\lfloor \ell/2 \rfloor$ and $\lfloor (\ell-1)/2 \rfloor$ of odd and even layers before $\ell$. The next eight coordinates are the entries of $\bx^{\odd}$ and $\bx^{\even}$. The final ten coordinates are used by $\bq$, in the order of the list above.
% : the range given by $B_1$ is for $y_t^2$, the ranges given by $B_2$, $B_3$, and $B_4$ are for the three quantities recorded for $R^{\odd}_{P_1,P_2}$ and again for the three quantities recorded for $R^{\even}_{P_1,P_2}$, and the range given by $B_5$ is for $t\cdot\bp[\bx^{\even}]_1$.

We include each edge prescribed by our update function~\eqref{eq:update} in $G$ whenever both endpoints belong to $V$, and we let $\calP$ consist of the critical paths starting at our source vertices $S$, one for each pair $(\ba,\bb) \in C^2$. It is straightforward to verify that every critical path $P \in \mathcal{P}$ we defined is fully contained in $G$, by our choice of the  coordinate ranges for the vertex set $V$.
% \ghnote{do we need additional justification here? it is very tedious to check all 20 coordinates lol.}

% Each step of a critical path increases the first three entries of $\bx^{\odd}$ or $\bx^{\even}$ by at most $q$ and the fourth entry by at most $q^2$, so a path starting in $S$ keeps the first three entries at most $X + Lq = 2X$ and the fourth entry at most $X' + Lq^2 = 2X'$. Then $y_t^2 < q^2$, $\bp[\bx^{\odd}]_1, \bp[\bx^{\even}]_1 \le 2Lq$, and $\bp[r_{\odd}], \bp[r_{\even}] \le L$. Hence each recorded quantity is nonnegative and at most $B_i/L$ for its range $B_i$: we get $q^2$ for $y_t^2$; $2Lq$, $2L^2q$, and $4L^2q^2$ for the three quantities recorded for $R^{\odd}_{P_1,P_2}$ and for $R^{\even}_{P_1,P_2}$; and $2Lq^2$ for $t\cdot\bp[\bx^{\even}]_1$. (As in \Cref{sec:non-optimized}, each $B_i$ is $L$ times the largest possible increment of its coordinate.) Thus each coordinate of $\bq$ stays at most $B_i + L\cdot B_i/L = 2B_i$. 
% Then every critical path we defined is contained in $G$.

We observe that
\[
 |S| = (X+1)^6\,(X'+1)^2\,(B_1+1)\,(B_2+1)^2\,(B_3+1)^2\,(B_4+1)^2\,(B_5+1)\,(B_6+1)\,(B_7+1) = \Theta(L^{55}),
\]
since $q = \Theta(L)$ gives $X = \Theta(L^2)$, $X' = \Theta(L^3)$, $B_1, B_2, B_6, B_7 = \Theta(L^3)$, $B_3, B_5 = \Theta(L^4)$, and $B_4 = \Theta(L^5)$. Each layer of $G$ has $\Theta(|S|)$ vertices, with an average of $\Theta(|C|) = \Theta(L)$ outgoing edges per vertex. Since every source and pair of directions define a distinct critical path, we obtain
\begin{equation}
 n = \Theta(L|S|) = \Theta(L^{56}), \qquad m = \Theta(L|S||C|) = \Theta(L^{57}), \qquad |\calP| = |S||C|^2 = \Theta(m). \label{eq:count}
\end{equation}
Our graph has only $|\calP| = \Theta(m)$ critical paths, while the graph of \Cref{sec:non-optimized} had $\Theta(mL)$. This difference comes from the number of directions: in both constructions, $|\calP| = |S||C|^2$ and $m = \Theta(L|S||C|)$, so $|\calP| = \Theta(m|C|/L)$, and \Cref{sec:non-optimized} used $|C| = L^2$ directions. We chose $|C| = \Theta(L)$ because this already gives $|\calP| = \Theta(m)$ critical paths, which is all that the counting argument below needs. More labels would increase the ranges of all coordinates without improving the bound. For comparison, in \Cref{sec:non-optimized}, the counters and sums of a source vertex contributed $30$ to the exponent of $n$, and its augmentation contributed $30030$. Here the sums contribute $18$ and the augmentation contributes $37$.

\paragraph{Verifying properties \ref{prop:unique}, \ref{prop:compatible-pairs}, and \ref{prop:small-overlap}.}
We showed in \Cref{subsec:optimized-splicing} that the direction set $C$ is average-free. The proofs of \Cref{clm:p1,clm:p2} do not depend on our choice of average-free set $C$, so they show that our critical paths satisfy~\ref{prop:unique} and~\ref{prop:small-overlap}, with $\rho = 1$. By \Cref{cla:splice}, they also satisfy the splicing property~\ref{prop:splicing-prop}, and hence~\ref{prop:compatible-pairs}.

\paragraph{Shortcutting subpaths of critical paths.}
In \Cref{sec:non-optimized}, we used a potential argument that rules out shortcuttings with at most $|\calP|/2$ edges, which was enough there because $|\calP| = \Theta(mL)$. Now $|\calP| = \Theta(m)$, so the same argument would rule out only shortcuttings with at most a small constant times $m$ edges. To handle shortcuttings with $O(m^{1+\eps})$ edges, we apply \Cref{thm:steiner-shortcut-template} to subpaths of the critical paths instead. If $G_+$ has diameter $D$, we split each critical path into subpaths of length $D+1$, and we show that each of these subpaths needs its own shortcut edge. We first check that families of such subpaths satisfy~\ref{prop:unique} and~\ref{prop:compatible-pairs}. Two compatible subpaths lie on compatible critical paths, but a priori, all common vertices of these critical paths might lie outside the two subpaths. To ensure~\ref{prop:compatible-pairs} for these subpaths, we therefore need the full splicing property~\ref{prop:splicing-prop}.
\begin{claim} \label{cla:subpaths}
Let $\calP'$ be a family of subpaths of critical paths in $\calP$. Then $\calP'$ satisfies~\ref{prop:unique} and~\ref{prop:compatible-pairs}. That is, each subpath in $\calP'$ is the unique path in $G$ between its endpoints, and any two compatible subpaths in $\calP'$ share a vertex.
\end{claim}
\begin{proof}
Every subpath of a critical path is the unique path between its endpoints, since otherwise the critical path would not be unique. This gives~\ref{prop:unique}. For~\ref{prop:compatible-pairs}, let $P[u,v]$ and $P'[u',v']$ be compatible subpaths in $\calP'$, where $P, P' \in \calP$. By compatibility, $G$ has a path from $u$ to $v'$ and a path from $u'$ to $v$. We can extend these paths to obtain cross-paths from $s(P)$ to $t(P')$ and from $s(P')$ to $t(P)$. Thus $P$ and $P'$ are compatible, and by \Cref{cla:splice}, the first of these cross-paths is $P[s(P), w] \circ P'[w, t(P')]$ for some vertex $w \in P \cap P'$. Hence $P$ and $P'$ share a vertex in a layer between $\lay(u)$ and $\lay(v')$. Indeed, if $\lay(w) < \lay(u)$, then the cross-path agrees with $P'$ after $w$ and passes through $u$, so $u \in P'$; if $\lay(w) > \lay(v')$, then similarly $v' \in P$; and otherwise, $w$ itself is such a vertex. In the same way, the second cross-path shows that $P$ and $P'$ share a vertex in a layer between $\lay(u')$ and $\lay(v)$.

By~\ref{prop:unique}, the layers of the common vertices of $P$ and $P'$ form an interval $I$, as in the proof of \Cref{lem:compatible-families-common}. We have shown that $I$ meets both $[\lay(u), \lay(v')]$ and $[\lay(u'), \lay(v)]$. Let $\ell_1 := \max\{\lay(u), \lay(u')\}$ and $\ell_2 := \min\{\lay(v), \lay(v')\}$. Then $\min I \le \ell_2$ and $\max I \ge \ell_1$. Since also $\ell_1 \le \ell_2$ (each of $u, u'$ can reach each of $v, v'$), the interval $I$ contains a layer between $\ell_1$ and $\ell_2$, and the common vertex of $P$ and $P'$ in this layer lies on both subpaths $P[u,v]$ and $P'[u',v']$.
\end{proof}

\Cref{thm:steiner-shortcut-template} holds for any family of paths of the same length that satisfies~\ref{prop:unique} and~\ref{prop:compatible-pairs}, since its proof in Appendix~\ref{sec:deferred-proof} does not use that the paths start in the first layer and end in the last layer of $G$. By \Cref{cla:subpaths}, we can then apply our Steiner shortcut lower bound template, \Cref{thm:steiner-shortcut-template}, to any family of subpaths of critical paths that all have the same length. We can now prove \Cref{thm:intro-shortcut-lb}.
\SteinerShortcutLowerBound*
\begin{proof}
Suppose that our lower bound graph $G$ admits a Steiner shortcutting $G_+$ with diameter $\diam(G_+) = D$. Split each critical path $P \in \calP$ into $\lfloor (L-1)/(D+1) \rfloor$ edge-disjoint subpaths with $D+1$ edges each, starting at $s(P)$, and discard the remaining edges of $P$. Let $\calP'$ be the family of these subpaths. The endpoints of each subpath in $\calP'$ are connected in $G$, so their distance in $G_+$ is at most $D$. By \Cref{cla:subpaths} and \Cref{thm:steiner-shortcut-template}, there is a (non-Steiner) shortcutting $G' = (V, E \cup E')$ with $|E'| \le |E_+|$ in which the endpoints of every subpath in $\calP'$ have distance at most $D$.

We will now prove that $D$ must be large, by charging each subpath in $\calP'$ to a distinct edge of $E'$. Fix a subpath $P[x,y] \in \calP'$ and a path in $G'$ from $x$ to $y$ with at most $D$ edges. Since $P[x, y]$ has $D+1$ edges, we know that there must be a shortcut edge $(u, v)$ on this $(x, y)$-path in $G'$ between two vertices at least two layers apart. We charge this edge $(u, v)$ to path $P[x, y]$. 
We never charge two distinct critical paths in $\mathcal{P}'$ to the same edge. Indeed, if the two paths are subpaths of the same path $P \in \calP$, they are edge-disjoint. Otherwise, by~\ref{prop:unique} both critical paths would then contain the path from $u$ to $v$, which has at least two edges; this contradicts~\ref{prop:small-overlap} with $\rho = 1$. Thus
\[
 |E_+| \ge |E'| \ge |\calP'| = |\calP| \left\lfloor \frac{L-1}{D+1} \right\rfloor.
\]
Since $|\calP| = \Theta(m)$ by~\eqref{eq:count}, we obtain
\[
 D = \Omega\left(\frac{L}{|E_+| + m} \cdot m\right) = \Omega\left(\frac{n^{1/56}}{|E_+| + m} \cdot m\right).
\]
Finally, if $|E_+| = O(m^{1+\eps})$, then $(|E_+| + m)/m = O(m^{\eps}) = O(n^{57\eps/56})$ by~\eqref{eq:count}, and hence $D \ge n^{1/56 - O(\eps)}$. This completes the proof of \Cref{thm:intro-shortcut-lb}.
\end{proof}

% \hfnoteinline{show some other test functions you can use to optimize a bit. higher priority optimizations just using test functions}

% \hfnoteinline{see below strategy for getting $n^{1/36}$:}
% \begin{enumerate}
%     \item First, relax (P3) to hitting sets. Simple lemma showing this is enough.
%     \item Second relax the framework of test functions to compatible families of paths.
%     \item Now use test functions to get:
%     \begin{enumerate}
%         \item Label confinement using a quadratic. This is actually written and
%         \item splicing on compatible families of three paths, but only one coordinate at a time (might not switch at the same time)
%         \item Show that the splice times for each of the two parity directions line up using another test function.
%         \item Without optimization, with gives $n^{1/46}$. Final optimization to get $n^{1/36}$ is to record everything mod a large prime.
%     \end{enumerate}
% \end{enumerate}
% \hfnoteinline{Can get $n^{1/36}$ with the following changes. First, relax (P3) to hitting sets. Second relax the framework of test functions to compatible families of paths. check label confinement with a degree $2$ polynomial, get splicing property on compatible families of size $3$, and }

\section{Extensions to Exact Hopsets and Monotone Circuits}

In this section, we extend our lower bounds on Steiner shortcuts to two new settings:
\begin{itemize}
    \item \textbf{Exact Hopsets.} We show polynomial hopbound lower bounds for exact hopsets with Steiner vertices in undirected unweighted graphs.
    \item \textbf{Monotone Circuits.} Instead of representing $G$ with a Steiner shortcut $G_+$, we consider the more general problem of representing $G$ by a monotone circuit with AND/OR gates. We prove polynomial depth lower bounds in this setting. 
\end{itemize}

\subsection{Extension to Exact Hopsets}
\label{sec:hopset-ext}

Here we prove our lower bounds for exact hopsets. Given a graph $G=(V, E, w)$, an exact hopset $H \subseteq V \times V$ with hopbound $\beta$ is a set of positively weighted edges such that:
\begin{itemize}
    \item for all $s, t \in V$, $\dist_{G \cup H}(s, t) = \dist_{G}(s, t)$, and
    \item for every pair of connected vertices $s, t \in V$, there exists a shortest $(s, t)$-path in $G \cup H$ with at most $\beta$ edges. 
\end{itemize}

Likewise, an exact Steiner hopset with hopbound $\beta$ is a graph $G_+ = (V \cup V_+, E \cup E_+, w_+)$, where $w_+$ is positive and agrees with $w$ on $E$, that satisfies the above two properties with $G_+$ in place of $G \cup H$. We prove the following lower bound for exact Steiner hopsets. 

\begin{corollary}
\label{cor:intro-hopset-lb}
There exists an infinite family of undirected graphs $G=(V,E)$ on $n$ nodes and $m$ edges such that any (exact) Steiner hopset $G_+ = (V \cup V_+, E \cup E_+, w)$ with $|E_+|=O(m^{1+\varepsilon})$ edges has hopbound at least
$n^{1/56 - O(\varepsilon)}$.
\end{corollary}

We will prove this corollary by a reduction to our Steiner shortcut lower bounds. The key idea is the following lemma, which allows us to translate exact Steiner hopsets into Steiner shortcuts in layered graphs.

\begin{lemma}
Let $\Vec{G}=(V, E)$ be a layered DAG, and let $G = (V, E')$ be its undirected version. If $G$ admits an exact Steiner hopset $H = (V \cup V_+, E' \cup E_+, w)$ with hopbound $\beta$, then $\Vec{G}$ admits a Steiner shortcut $\Vec{H}$ with at most $|E_+|$ added edges such that $\dist_{\Vec{H}}(x, y) \le \beta$ for all $x, y \in V$ such that $x$ can reach $y$ in $\Vec{G}$. 
\label{lem:reduction}
\end{lemma}
\begin{proof}
Let $H=(V \cup V_+, E' \cup E_+, w)$ be an exact Steiner hopset  of $G$. Let $L$ denote the number of layers in graphs $\Vec{G}$ and $G$, and let $\ell(v) \in [L]$ denote the layer of each vertex $v \in V$.  We will use the following basic fact about the layering function $\ell(\cdot)$:
\[
|\ell(x) - \ell(y)| \le \text{dist}_{G}(x, y)=\text{dist}_H(x, y)  \qquad\text{for all original vertices }x,y.
\]

We can extend the layering function $\ell(\cdot)$  to the vertices of $V_+$ as follows:
\[\ell(v) = \min_{u \in V}(\ell(u) + \text{dist}_H(u, v)). \]
We may assume that every vertex of $V_+$ is connected to $V$ in $H$, since the other vertices of $V_+$ can be deleted with their edges; then $\ell(v)$ is finite for all $v$. It is straightforward to verify that this new layering function $\ell(\cdot)$ is consistent with the original layering function on vertices $v \in V$.

We are now ready to define our Steiner shortcut $\Vec{H}$. Add all vertices in $V \cup V_+$ to $\Vec{H}$. For each edge $\{u, v\} \in E' \cup E_+$:
\begin{itemize}
    \item If $\ell(v) - \ell(u) = w(u, v)$, then add the directed edge $(u, v)$ to $\Vec{H}$.
    \item If $\ell(u) - \ell(v) = w(u, v)$, then add the directed edge $(v, u)$ to $\Vec{H}$.
\end{itemize}
This completes the construction of $\Vec{H}$. 
Each edge $\{u, v\} \in E'$ has weight $1$ and joins consecutive layers, so it contributes exactly its directed edge in $E$. Each edge in $E_+$ has positive weight, so it contributes at most one directed edge. Hence $\Vec{H}$ consists of $\Vec{G}$ and at most $|E_+|$ added edges.
What remains is to show that $\Vec{H}$ is a valid Steiner shortcut of $\Vec{G}$ in which $\dist_{\Vec{H}}(x, y) \le \beta$ for all $x, y \in V$ such that $x$ can reach $y$ in $\Vec{G}$.
We begin by showing that $\Vec{H}$ does not introduce new reachability between original vertices in $V$. 

Consider vertices $x, y \in V$ such that $x$ can reach $y$ in $\Vec{H}$, and let $P=(x=v_1, v_2, \dots, v_r=y)$ be an $(x, y)$-path in $\Vec{H}$. Then by our construction of $\Vec{H}$ and a telescoping argument,
\[
 w(P)  = \sum_{i=1}^{r-1} (\ell(v_{i+1}) - \ell(v_i)) = 
\ell(y) - \ell(x).
\]
We conclude that $\text{dist}_G(x, y) \le \ell(y) - \ell(x)$, since $\text{dist}_G(x, y) = \text{dist}_H(x, y) \le w(P)$. Then there exists an $(x, y)$-path in $G$ that uses exactly one edge in each layer between $x$ and $y$. This path must also be a directed path in $\Vec{G}$, so  $x$ can reach $y$ in $\Vec{G}$ as claimed.

Now we show that $\dist_{\Vec{H}}(x, y) \le \beta$ for every such pair. Suppose that $x$ can reach $y$ in $\Vec{G}$. Then $\text{dist}_G(x, y) = \ell(y) - \ell(x)$. Let $P$ be an $(x, y)$-path in $H$ with weight $\ell(y) - \ell(x)$ and at most $\beta$ edges. Then 
\[
\sum_{(u, v) \in P } (\ell(v) - \ell(u))
= \ell(y) - \ell(x) = \sum_{(u, v) \in P } w(u, v). 
\]
By the triangle inequality, $\ell(v) - \ell(u) \le w(u, v)$ for every $\{u, v\}  \in  H$, so we conclude that $w(u, v) = \ell(v) - \ell(u)$ for every $(u, v) \in P$. Path $P$ survives as an $(x, y)$-path in $\Vec{H}$ with at most $\beta$ edges.
\end{proof}

\Cref{lem:reduction} allows us to translate layered Steiner shortcut lower bound graphs into undirected exact hopset lower bound graphs. The only property we need from our Steiner shortcut set construction is that it is layered. The lower bound graphs of \Cref{thm:intro-shortcut-lb} are layered, so we directly obtain \Cref{cor:intro-hopset-lb}.

\begin{proof}[Proof of \Cref{cor:intro-hopset-lb}]
Let $\Vec{G}$ be a lower bound graph of \Cref{thm:intro-shortcut-lb} with $n$ nodes and $m$ edges, and let $G$ be its undirected version, which also has $n$ nodes and $m$ edges. Suppose that $G$ admits an exact Steiner hopset with $|E_+| = O(m^{1+\eps})$ edges and hopbound $\beta$. By \Cref{lem:reduction}, $\Vec{G}$ admits a Steiner shortcut $G_+$ with at most $|E_+|$ added edges such that $\dist_{G_+}(x, y) \le \beta$ for all $x, y \in V$ such that $x$ can reach $y$ in $\Vec{G}$. The proof of \Cref{thm:intro-shortcut-lb} bounds distances in $G_+$ only between such pairs (the endpoints of subpaths of critical paths), so it gives $\beta \ge n^{1/56 - O(\eps)}$.
\end{proof}

\subsection{Extension to Monotone Circuits}
\label{sec:monotone-ckt-ext} 

In this subsection, we extend our Steiner shortcut lower bounds to obtain monotone circuit lower bounds. Before we can formally state and prove this corollary, we need to introduce some background.

We begin by stating the circuit model and notation. Let $A$ be
a Boolean $n\times n$ matrix. A circuit $C$ computing $A$ receives an input $\bx\in\{0,1\}^n$
and produces an output $\by\in\{0,1\}^n$ such that $\by=A\bx$,
where addition is Boolean OR and multiplication is Boolean AND.
Explicitly,
\[
  y_i=\bigvee_{j=1}^n (A_{i,j}\wedge x_j).
\]
Here $x_j$ is the value at input node $j$, and $y_i$ is the value
at output node $i$. Thus row $i$ of $A$ specifies which inputs
are included in the OR computed by output $i$.

We say that $C$ is an OR circuit if it contains only OR gates. A circuit is monotone if it uses only OR and AND gates. Circuit size is the number of wires, and depth is the maximum number of wires on a directed path from an input node to an output node. Following the notation  of~\cite{JuknaSergeev2013}, we write $\operatorname{OR}(A)$ for the minimum number of wires in an OR circuit computing $A$. With this background, we are ready to state the result we prove in this section. 

\begin{corollary}
\label{cor:intro-monotone-circuit-lb}
There exists an infinite family of Boolean matrices
$A\in\{0,1\}^{n\times n}$ such that any monotone AND/OR circuit $C$ computing $A$ with $O(\text{\normalfont OR}(A)^{1+\varepsilon})$ wires has
\(
  \operatorname{depth}(C)\ge n^{1/56-O(\varepsilon)}.
\)
\end{corollary}

For an OR circuit computing $A$, the entry $A_{i,j}$ is one exactly
when there is a directed circuit path from input $j$ to output $i$.
In monotone circuits with OR and AND gates, a path from input $j$ to output $i$ can exist  even if $A_{i,j}=0$. Our first lemma shows that we can nevertheless choose paths for each of the one-entries of $A$ with a useful guarantee on their intersections.

\begin{lemma}[Tracing paths in monotone circuits]
\label{lem:monotone-tracing}
Let $C$ be a monotone AND/OR circuit computing a Boolean matrix $A$. 
Then there exists a collection $\mathcal{P}$ of directed paths in $C$ such that:
\begin{enumerate}
    \item For each $(i, j) \in [n]^2$ with $A_{i, j} = 1$, the collection $\mathcal{P}$ contains a path $P_{i, j}$ from input $j$ to output $i$,
    \item For any two paths $P_{i, j}, P_{i', j'} \in \mathcal{P}$, if these paths intersect at a common node (gate), then $A_{i, j'} = A_{i', j} = 1$, and
    \item Each path $P_{i, j} \in \mathcal{P}$ has length at most $\operatorname{depth}(C)$.
\end{enumerate}
\end{lemma}

\Cref{lem:monotone-tracing} is trivially true for OR circuits. However, for monotone circuits, item 2 is not immediate. We prove this lemma using a path-tracing argument due to Karchmer and Wigderson~\cite{KW90}.

\begin{proof}
For each $j\in[n]$, let $\be_j\in\{0,1\}^n$ be the vector whose
only one-entry is in coordinate $j$. For each $i\in[n]$, let
$\bz^{(i)}\in\{0,1\}^n$ be defined by
\[
 z^{(i)}_k=1-A_{i,k}.
\]
The value of output gate $i$ on input $\bz^{(i)}$ is zero.  Write $g(\bx)$
for the value of a node $g$ (gate) on input $\bx$.

Fix $(i,j)$ with $A_{i,j}=1$. Starting at output gate $i$, we trace
backwards through the circuit, maintaining
\[
 g(\be_j)=1
 \qquad\text{and}\qquad
 g(\bz^{(i)})=0
\]
at every visited node $g$.  These conditions must hold  at output $i$. Let $g$ be the current node we are visiting in our trace.
\begin{itemize}
    \item Suppose that $g$ is an OR gate. Then we choose a predecessor gate $g'$ with $g'(\be_j) = 1$. This gate exists since $g(\be_j) = 1$ and $g$ is an OR gate. Moreover, $g'(\bz^{(i)}) = 0$, since $g(\bz^{(i)}) = 0$.
    \item Suppose that $g$ is an AND gate. Then we choose a predecessor gate $g'$ with $g'(\bz^{(i)}) = 0$. This gate exists since $g(\bz^{(i)}) = 0$ and $g$ is an AND gate. Moreover, $g'(\be_j) = 1$, since $g(\be_j) = 1$.
\end{itemize}
Thus both conditions are preserved. The trace ends at an input,
which must be input gate $j$, since this is the only input gate that is one on input $\be_j$. Reversing this trace gives us the claimed $(j, i)$-path $P_{i,j}$.

Let $\mathcal{P}$ consist of these paths, and suppose that
$P_{i,j}$ and $P_{i',j'}$ share a node $g$. The tracing
conditions give
\[
 g(\be_{j'})=1
 \qquad\text{and}\qquad
 g(\bz^{(i)})=0.
\]
If $A_{i,j'}=0$, then $\be_{j'}\le \bz^{(i)}$ coordinatewise. However, at gate $g$, $g(\be_{j'})=1$ and $g(\bz^{(i)})=0$, contradicting the monotonicity of circuit $C$. We conclude that $A_{i,j'}=1$.
A symmetric argument shows $A_{i',j}=1$ as well. Finally, each path $P_{i,j}$ is a directed input-to-output path in
$C$, so its length is at most $\operatorname{depth}(C)$.
\end{proof}

We now prove \Cref{cor:intro-monotone-circuit-lb}. The circuit paths of \Cref{lem:monotone-tracing} play the role of the shortcut paths $P_+$, and the nodes of the circuit play the role of the Steiner vertices. As in the proof of \Cref{thm:intro-shortcut-lb}, we work with subpaths of critical paths.

\begin{proof}[Proof of \Cref{cor:intro-monotone-circuit-lb}]
Let $G = (V,E)$ be a lower bound graph of \Cref{thm:intro-shortcut-lb} with $n$ nodes and $m$ edges, and let $\mathcal{P}$ be our set of critical paths. Our matrix $A$ is the reachability matrix of $G$. The  rows and columns of $A$ are indexed by $V$, and $A_{u,v} = 1$ exactly when $v$ can reach $u$ in $G$. In particular, $A_{v, v}=1$ for all $v \in V$. Since $A$ is indexed by $V$, we also index the inputs and outputs of any circuit computing $A$ by $V$. For each $v \in V$, let $\mathrm{in}_v$ denote the input gate that receives $x_v$, and let $\mathrm{out}_v$ denote the output gate that computes $y_v$.

We first show that $\operatorname{OR}(A) = O(m)$. Consider the following OR circuit constructed from $G$. Add an OR gate $g_v$ to the circuit for each vertex $v \in V$. Add a wire from $\mathrm{in}_v$ to $g_v$ and a wire from $g_v$ to $\mathrm{out}_v$. Additionally, for each edge $(u, v) \in E$ add a wire from $g_u$ to $g_v$. It is straightforward to verify that this OR circuit computes $A$. It has $2n + m$ wires, and hence $\operatorname{OR}(A) \le 2n + m = O(m)$ by~\eqref{eq:count}.

Now let $C$ be a monotone circuit computing $A$ with depth $d$. Split each critical path $P \in \calP$ into $\lfloor (L-1)/(d+1) \rfloor$ edge-disjoint subpaths with $d+1$ edges each, starting at $s(P)$, and let $\calP'$ be the family of these subpaths. For each subpath $P' \in \calP'$, the vertex $s(P')$ can reach $t(P')$, so $A_{t(P'), s(P')} = 1$. Let $T_{P'}$ be the path $P_{t(P'), s(P')}$ in $C$ given by \Cref{lem:monotone-tracing}. It goes from $\mathrm{in}_{s(P')}$ to $\mathrm{out}_{t(P')}$ and has at most $d$ wires.

For each gate $g$ of $C$, let $\calF_g$ be the set of subpaths $P' \in \calP'$ such that $T_{P'}$ contains $g$. By the second property of \Cref{lem:monotone-tracing}, for all $P', Q' \in \calF_g$, we have $A_{t(P'), s(Q')} = 1$, that is, $s(Q')$ can reach $t(P')$ in $G$. Hence $\calF_g$ is a compatible family. By \Cref{cla:subpaths}, the family $\calP'$ satisfies~\ref{prop:unique} and~\ref{prop:compatible-pairs}. The proof of \Cref{lem:compatible-families-common} uses only these two properties, so whenever $\calF_g$ is nonempty, its subpaths have a common vertex in $G$. We choose such a common vertex $v_g \in V(G)$ as follows.
\begin{itemize}
 \item If $g = \mathrm{in}_v$ or $g = \mathrm{out}_v$ for some $v \in V$, we let $v_g := v$. This is a common vertex of $\calF_g$. Indeed, since $A_{v,v} = 1$, \Cref{lem:monotone-tracing} gives a path $P_{v,v}$ in $C$ from $\mathrm{in}_v$ to $\mathrm{out}_v$, which contains $g$. For each $P' \in \calF_g$, the path $T_{P'} = P_{t(P'), s(P')}$ also contains $g$, so the second property of \Cref{lem:monotone-tracing} gives $A_{v, s(P')} = A_{t(P'), v} = 1$. That is, $s(P')$ can reach $v$, and $v$ can reach $t(P')$. Since $P'$ is the unique path from $s(P')$ to $t(P')$ by~\ref{prop:unique}, it contains $v$.
 \item Otherwise, we let $v_g$ be any common vertex of $\calF_g$. 
\end{itemize}

We now charge each subpath in $\calP'$ to a distinct wire of $C$. Fix $P' \in \calP'$, and let $g_0, g_1, \ldots, g_k$ be the gates of $T_{P'}$, where $g_0 = \mathrm{in}_{s(P')}$, $g_k = \mathrm{out}_{t(P')}$, and $k \le d$. Since $P' \in \calF_{g_r}$ for every $r$, the vertices $v_{g_0} = s(P'), v_{g_1}, \ldots, v_{g_k} = t(P')$ all lie on $P'$. Their layers go from $\lay(s(P'))$ to $\lay(t(P')) = \lay(s(P')) + d + 1$ in $k \le d$ steps.\footnote{Recall that we use $\lay(v)$ to denote the layer of vertex $v$ in the lower bound graph $G$.} Hence some wire $(g_{r-1}, g_r)$ of $T_{P'}$ satisfies $\lay(v_{g_r}) - \lay(v_{g_{r-1}}) \ge 2$, and we charge $P'$ to this wire.

Two subpaths are never charged to the same wire. Indeed, suppose that subpaths $P'_1 \subseteq P_1$ and $P'_2 \subseteq P_2$ in $\calP'$ are both charged to a wire $(g, g')$, where $P_1, P_2 \in \calP$. Then both subpaths belong to $\calF_g$ and $\calF_{g'}$, so both contain $v_g$ and $v_{g'}$, and $\lay(v_{g'}) - \lay(v_g) \ge 2$. By~\ref{prop:unique}, both $P_1$ and $P_2$ contain the path in $G$ from $v_g$ to $v_{g'}$, which has at least two edges. If $P_1 \ne P_2$, this contradicts~\ref{prop:small-overlap} with $\rho = 1$. If $P_1 = P_2$, it contradicts the fact that distinct subpaths of the same critical path are edge-disjoint. Thus
\[
 \#\text{wires of } C \ \ge\ |\calP'| = |\calP| \left\lfloor \frac{L-1}{d+1} \right\rfloor.
\]
Finally, suppose that $C$ has $O(\operatorname{OR}(A)^{1+\eps}) = O(m^{1+\eps})$ wires. Since $|\calP| = \Theta(m)$ by~\eqref{eq:count}, the same calculation as at the end of the proof of \Cref{thm:intro-shortcut-lb}, with $d$ in place of $D$ and the number of wires of $C$ in place of $|E_+|$, gives $\operatorname{depth}(C) = d \ge n^{1/56 - O(\eps)}$, where $n = |V|$ is the dimension of $A$.
\end{proof}

\section*{AI Disclosure}
% The main result, a refutation of \Cref{conjecture:steiner}, was disovered by ChatGPT 6 Astra with very limited interaction. The idea to use test functions to ensure properties of cross-paths of compatible critical paths was present, but completely obfuscated, in the original AI-produced manuscript. We then, in concert with various AI models, iteratively simplified the result and extracted the main insights. After our work in isolating the test function framework, the choice of test functions yielding the optimized $n^{1/56}$ diameter lower bound (discussed in \Cref{sec:optimized}) was discovered by ChatGPT 6 Astra. We note that previous attempts to optimize the lower bound had yielded manuscripts claiming even stronger lower bounds, as high as $n^{1/26}$, but these both did not fit nicely into our framework nor admit obvious transparent explanations. The ideas to consider both extensions to exact Steiner hopsets and monotone circuits were human, but the proof of the latter was initially AI-generated. Various AI models were used throughout for editorial aspects. We take full responsibility for the manuscript's contents and the correctness of its results.

% \aaroninline{My new attempt below}
Our main result, a refutation of \Cref{conjecture:steiner}, was  discovered by ChatGPT 6 Astra with very limited interaction. That initial proof was significantly more complicated and less modular: it proved that a certain elaborate function satisfies specific properties, and there was no clear general technique. 
% We then, in concert with various AI models,  went through many iterations of simplifying the result and distilling the main insights.
We then went through many iterations of simplifying the result and distilling the main insights. This was done mainly through discussions among the authors, with AI models assisting along the way.

We eventually succeeded in developing our test function framework, which gives a significantly clearer and more unified construction method.  This framework led to our short and simple presentation of the non-optimized bound in \Cref{sec:non-optimized} (\Cref{thm:st-shortcut-lb}), which was written entirely from scratch by us.

For the more optimized $n^{1/56}$ lower bound (\Cref{sec:optimized}, \Cref{thm:intro-shortcut-lb}), we prompted ChatGPT Astra with our test function framework, and the specific choice of test functions we use was discovered by Astra. This section is mostly written by us from scratch, although AI wrote the initial draft of a few technical components, which we then significantly edited.

We note that previous attempts to optimize the lower bound had yielded ChatGPT manuscripts claiming even
stronger lower bounds, as high as $n^{1/26}$, but these neither fit nicely into our framework nor
admit obvious transparent explanations. The ideas to consider both extensions to exact Steiner
hopsets (\Cref{sec:hopset-ext}) and monotone circuits (\Cref{sec:monotone-ckt-ext}) were human, but the proof of the latter (\Cref{cor:intro-monotone-circuit-lb}) was initially AI-generated.
Various AI models were used throughout for editorial aspects. We take full responsibility for the
manuscript's contents and the correctness of its results.

\bibliographystyle{alpha}
\bibliography{refs}

\appendix 

\section{Proof of \Cref{thm:steiner-shortcut-template}} \label{sec:deferred-proof}
In this section we complete the proof of \Cref{thm:steiner-shortcut-template} by generalizing the sketch given in \Cref{sec:sufficient-props}. We restate the theorem for convenience.
\lbtemplate*
Also recall the following useful lemma we proved about compatible families of critical paths.
\compatcommon*
The candidate approach is then to replace each Steiner point $x$ with the vertex in $V$ contained in its corresponding compatible family $\calF_x$. That is exactly what we will do, but with some care to ensure that the transformation preserves the reachability relation of the original graph.
\begin{proof}[Proof of \Cref{thm:steiner-shortcut-template}]
Let $G_+$ be a Steiner shortcutting of $G$. Fix a shortest path $P_+$ for each $P\in\calP$, and discard added
edges not used by any of these paths. Every original vertex on
$P_+$ belongs to $P$, by uniqueness of critical paths \ref{prop:unique} and the fact that Steiner shortcuts preserve reachability.

We replace each connected component of Steiner vertices all at
once, where connectivity ignores edge directions. Fix such a
component, and let $H$ consist of its vertices, its incident edges,
and the original vertex endpoints of the incident edges; see the left-hand side of \Cref{fig:steiner-component-map}. 

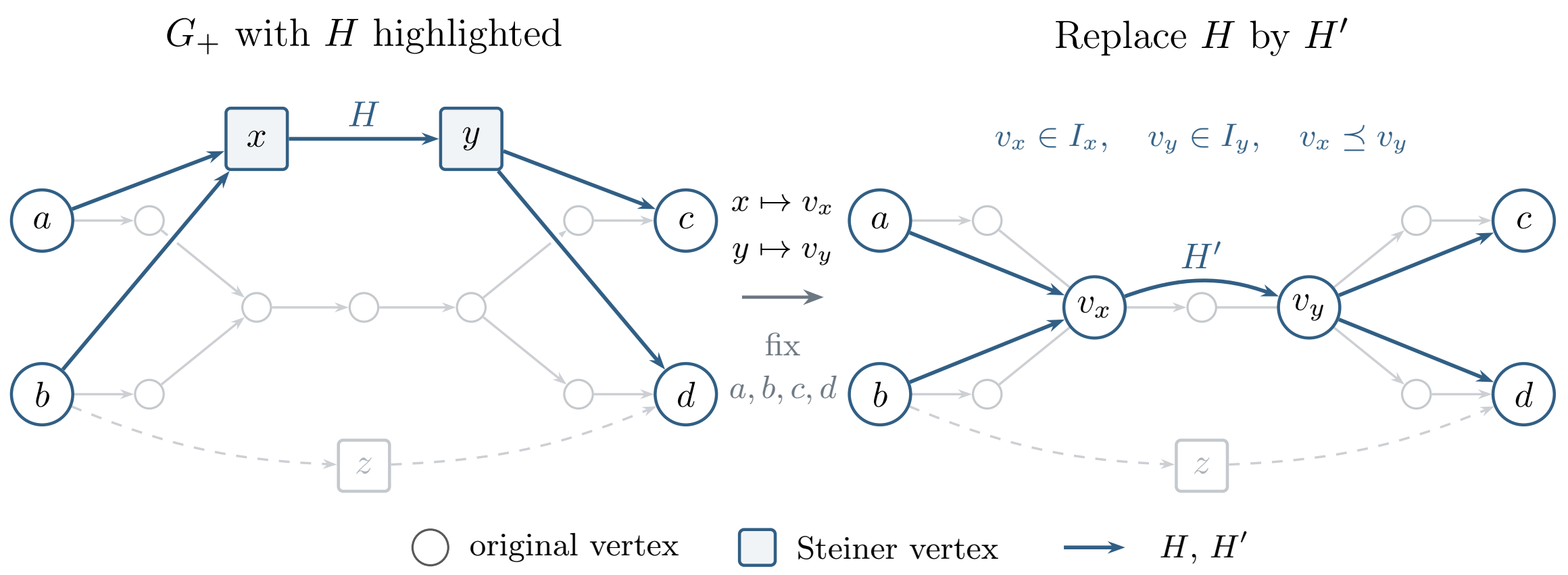
\begin{figure}[tbp]
    \centering
    \begingroup
\definecolor{componentblue}{HTML}{315F85}
\definecolor{componentgray}{HTML}{6C7782}
\begin{tikzpicture}[
    font=\small,
    >={Stealth[length=1.7mm,width=1.2mm]},
    original/.style={circle,draw=componentgray!40,fill=white,
        line width=.6pt,minimum size=2.8mm,inner sep=0pt},
    boundary/.style={original,minimum size=6mm},
    active original/.style={boundary,draw=componentblue,
        line width=.85pt,text=black},
    steiner/.style={rectangle,rounded corners=1.3pt,
        draw=componentblue,fill=componentblue!7,line width=.85pt,
        minimum size=6mm,inner sep=0pt},
    background edge/.style={->,draw=componentgray!35,line width=.65pt,
        >={Stealth[length=1.35mm,width=.95mm]}},
    background shortcut/.style={background edge,dashed},
    component edge/.style={->,draw=componentblue,line width=1.05pt,
        preaction={draw=white,line width=3.4pt,-,
            shorten <=1pt,shorten >=1pt}},
]

% The same original vertices and edges appear at the same positions
% in both panels.  Small circles are original vertices outside H or H'.
\foreach \shift in {0,8.2} {
    \begin{scope}[xshift=\shift cm]
        \node[boundary] (a) at (0,.85) {};
        \node[boundary] (b) at (0,-.85) {};
        \node[original] (aone) at (1.05,.85) {};
        \node[original] (bone) at (1.05,-.85) {};
        \node[original] (vx) at (2.1,0) {};
        \node[original] (middle) at (3.15,0) {};
        \node[original] (vy) at (4.2,0) {};
        \node[original] (cone) at (5.25,.85) {};
        \node[original] (done) at (5.25,-.85) {};
        \node[boundary] (c) at (6.3,.85) {};
        \node[boundary] (d) at (6.3,-.85) {};
        \foreach \u/\v in {a/aone,b/bone,aone/vx,bone/vx,
            vx/middle,middle/vy,vy/cone,vy/done,cone/c,done/d} {
            \draw[background edge] (\u) -- (\v);
        }

        % A second component of Steiner vertices, outside the selected H.
        \node[steiner,draw=componentgray!40,fill=white,
            text=componentgray!55,minimum size=5mm] (z) at (3.15,-1.55) {$z$};
        \draw[background shortcut] (b) to[bend right=10] (z);
        \draw[background shortcut] (z) to[bend right=10] (d);

        % Original boundary vertices stay fixed under the replacement.
        \node[active original] (a) at (a) {$a$};
        \node[active original] (b) at (b) {$b$};
        \node[active original] (c) at (c) {$c$};
        \node[active original] (d) at (d) {$d$};

        \ifdim\shift pt=0pt
            \node[font=\normalsize] at (3.15,2.65) {$G_+$ with $H$ highlighted};
            \node[steiner] (x) at (2.1,1.65) {$x$};
            \node[steiner] (y) at (4.2,1.65) {$y$};
            \draw[component edge] (a) -- (x);
            \draw[component edge] (b) -- (x);
            \draw[component edge] (x) --
                node[above,text=componentblue,inner sep=3pt] {$H$} (y);
            \draw[component edge] (y) -- (c);
            \draw[component edge] (y) -- (d);
        \else
            \node[font=\normalsize] at (3.15,2.65) {Replace $H$ by $H'$};
            \node[active original] (vx) at (vx) {$v_x$};
            \node[active original] (vy) at (vy) {$v_y$};
            \draw[component edge] (a) -- (vx);
            \draw[component edge] (b) -- (vx);
            \draw[component edge] (vx) to[bend left=20]
                node[above,text=componentblue,inner sep=3pt] {$H'$} (vy);
            \draw[component edge] (vy) -- (c);
            \draw[component edge] (vy) -- (d);
            \node[font=\footnotesize,text=componentblue] at (3.15,1.65)
                {$v_x\in I_x,\quad v_y\in I_y,\quad v_x\preceq v_y$};
        \fi
    \end{scope}
}

% A single map between the panels; the original boundary is fixed.
\draw[->,draw=componentgray,line width=.9pt,
    >={Stealth[length=2mm,width=1.5mm]}] (6.85,.1) -- (7.65,.1);
\node[font=\footnotesize,align=center] at (7.25,.75)
    {$x\mapsto v_x$\\[2pt]$y\mapsto v_y$};
\node[font=\footnotesize,text=componentgray,align=center] at (7.25,-.6)
    {fix\\[1pt]$a,b,c,d$};

\node[original,draw=black!65,minimum size=3.5mm] at (3.8,-2.35) {};
\node[anchor=west,font=\footnotesize] at (4.05,-2.35) {original vertex};
\node[steiner,minimum size=3.5mm] at (7.0,-2.35) {};
\node[anchor=west,font=\footnotesize] at (7.25,-2.35) {Steiner vertex};
\draw[component edge] (10,-2.35) -- (10.6,-2.35);
\node[anchor=west,font=\footnotesize] at (10.8,-2.35) {$H$, $H'$};
\end{tikzpicture}
\endgroup
    \caption{The replacement of a connected component of Steiner vertices $H$ (blue) by a shortcutting $H'$ on the original
    vertices. The map fixes original vertices $a,b,c,d$ and
    maps Steiner vertices $x,y$ to $v_x\in I_x$, $v_y\in I_y$.}
    \label{fig:steiner-component-map}
\end{figure}

For each vertex $x$ in $H$, let $\calF_x=\{P\in\calP:x\in P_+\}$ be the corresponding compatible family. Furthermore, for a Steiner vertex $x$, let $I_x$ be the set of common vertices of
$\calF_x$. These are the candidate replacements for $x$. For an original vertex $x$, set $I_x=\{x\}$: we simply
require original vertices to stay fixed. In either case, $I_x$
is a nonempty interval $[\ell_x,r_x]$ along every member of $\calF_x$.

We will replace $H$ with an identical copy $H'$ instead supported on the original vertex set; each vertex $x$ in $H$ will have a corresponding $v_x \in I_x$ such that $(x,y) \in E(H) \iff (v_x, v_y) \in E(H')$. See \Cref{fig:steiner-component-map}. Suppose for a moment that such a replacement were valid, i.e., it preserved reachability. Then replacing $H$ by $H'$ retains the same number of shortcut edges. Moreover, for each subpath of $P_+$ for $P \in \calP$ passing through $H$, it is mapped to a subpath of the same length via the map between the edges of $H$ and $H'$. This remains a shortcutting of $P$ because of the fact that $v_x = x$ for each original vertex $x$ in $H$. The result then follows by replacing each connected component $H$ of Steiner vertices in this way. 

So, it remains to show that $H'$ preserves reachability. For this, we need to define $v_z$'s such that $v_x \preceq v_y$ for each $(x,y) \in E(H)$. We use the following definition:
\[
\mu_y:=\max\{\ell_u:u\text{ can reach }y\text{ in }H\}.
\]
We choose $v_y \in I_y$ at layer $\mu_y$. This is promising---note that $\mu_x\le \mu_y$ for every
edge $(x,y)$ of $H$, since every node that can reach $x$ can also reach $y$. As such, if this were a valid way to select $v_y$'s, it would immediately imply that $H'$ preserves reachability. 

So, the last remaining detail is to verify that there actually \emph{exists} such a choice of $v_y \in I_y$ at layer $\mu_y$, i.e., $\mu_y \in [\ell_y, r_y]$. We trivially have that $\ell_y \leq \mu_y$ since $y$ can reach itself. The other bound, $\mu_y \leq r_y$, also holds, as we now show.
\begin{claim} \label{cla:st-to-non-st-sub}
For each vertex $y$ of $H$, $\mu_y \leq r_y$.
\end{claim}
\begin{proof}[Proof of \Cref{cla:st-to-non-st-sub}]
\renewcommand{\qedsymbol}{\color{black!65}\(\triangle\)}
Choose an ancestor $u$ of $y$ with $\ell_u=\mu_y$, and let
$w_u$ be the first vertex of $I_u$. Assume for the sake of contradiction that $w_u\notin I_y$.
Along a path from $u$ to $y$, let $x$ be the first vertex
with $w_u\notin I_x$, with $v$ its predecessor along the path.
\begin{figure}[tbp]
    \centering
    \begingroup
\definecolor{pathblue}{HTML}{315F85}
\definecolor{edgegray}{HTML}{6C7782}
\begin{tikzpicture}[
    font=\small,
    >={Stealth[length=1.7mm,width=1.2mm]},
    vertex/.style={circle,draw=black!70,fill=white,line width=.65pt,
        minimum size=6mm,inner sep=0pt},
    endpoint/.style={vertex,minimum size=3mm},
    selected vertex/.style={vertex,draw=pathblue,fill=pathblue!8,
        double,double distance=.6pt,minimum size=7mm},
    critical path/.style={->,draw=pathblue,line width=1.05pt},
    component path/.style={->,draw=edgegray,line width=.9pt,dashed},
    path label/.style={fill=white,inner sep=2pt,font=\small},
]

% The selected path in H and a shortcutted critical path share (v,x).
\node[font=\normalsize] at (7.4,2.95)
    {In $G_+$: the path from $u$ to $y$ in $H$ and $P^*_+$};
\node[vertex] (u) at (1.8,.65) {$u$};
\node[vertex] (v) at (5.1,.65) {$v$};
\node[vertex] (x) at (9.7,.65) {$x$};
\node[vertex] (y) at (13,.65) {$y$};
\node[endpoint] (splus) at (1.8,2) {};
\node[endpoint] (tplus) at (13,2) {};
\node[above=3pt] at (splus.north) {$s(P^*)$};
\node[above=3pt] at (tplus.north) {$t(P^*)$};

\draw[component path] (u) -- (v);
\draw[critical path] (v) --
    node[path label,above=4pt,text=pathblue] {$P^*_+$} (x);
\draw[component path] (x) -- (y);
\draw[critical path,dashed] (splus) to[out=0,in=150] (v);
\draw[critical path,dashed] (x) to[out=30,in=180] (tplus);

\node[font=\footnotesize] at (1.8,-.05) {$\ell_u=\mu_y$};
\node[font=\footnotesize] at (5.1,-.05) {$w_u\in I_v$};
\node[font=\footnotesize] at (9.7,-.05) {$w_u\notin I_x$};
\draw[black!15,line width=.5pt] (.5,-.55) -- (14.3,-.55);

% On P^*, the whole interval I_x lies before w_u. The node w_u is
% known to be on P^* because w_u lies in I_v and P^* belongs to F_v.
\node[font=\normalsize] at (7.4,-1.05)
    {In $G$: the corresponding critical path $P^*$};
\draw[draw=pathblue!18,line width=7pt,line cap=round]
    (3.6,-2.25) -- (6.6,-2.25);
\node[endpoint] (s) at (1.8,-2.25) {};
\node[vertex,minimum size=7mm] (wx) at (3.6,-2.25) {$w_x$};
\node[endpoint] (middle) at (5.1,-2.25) {};
\node[endpoint] (rightx) at (6.6,-2.25) {};
\node[selected vertex] (wu) at (9.7,-2.25) {$w_u$};
\node[endpoint] (t) at (13,-2.25) {};

\draw[critical path,dashed] (s) -- (wx);
\draw[critical path] (wx) -- (middle);
\draw[critical path] (middle) -- (rightx);
\draw[critical path,dashed] (rightx) -- (wu);
\draw[critical path,dashed] (wu) -- (t);
\node[below=4pt] at (s.south) {$s(P^*)$};
\node[below=4pt] at (t.south) {$t(P^*)$};

\draw[draw=pathblue,line width=.65pt]
    (3.6,-1.9) -- (3.6,-1.7) -- (6.6,-1.7) -- (6.6,-1.9);
\node[text=pathblue,fill=white,inner sep=2pt] at (5.1,-1.7) {$I_x$};
\node[font=\footnotesize] at (3.6,-2.95) {$\ell_x$};
\node[font=\footnotesize] at (6.6,-2.95) {$r_x$};
\node[font=\footnotesize] at (9.7,-2.95) {$\ell_u=\mu_y$};
\node at (7.4,-3.55) {$\ell_x\le r_x<\ell_u=\mu_y$};
\end{tikzpicture}
\endgroup
    \caption{The hypothetical configuration in the proof of
    \Cref{cla:st-to-non-st-sub}. Top: $x$ is the first vertex on the
    $u$--$y$ path whose interval excludes $w_u$, and $P^*_+$ contains
    the edge $(v,x)$. 
    Bottom: along $P^*$, the interval $I_x$ lies strictly before
    $w_u\in I_v$; $w_x$ is the first vertex of $I_x$.}
    \label{fig:interval-order-claim}
\end{figure}
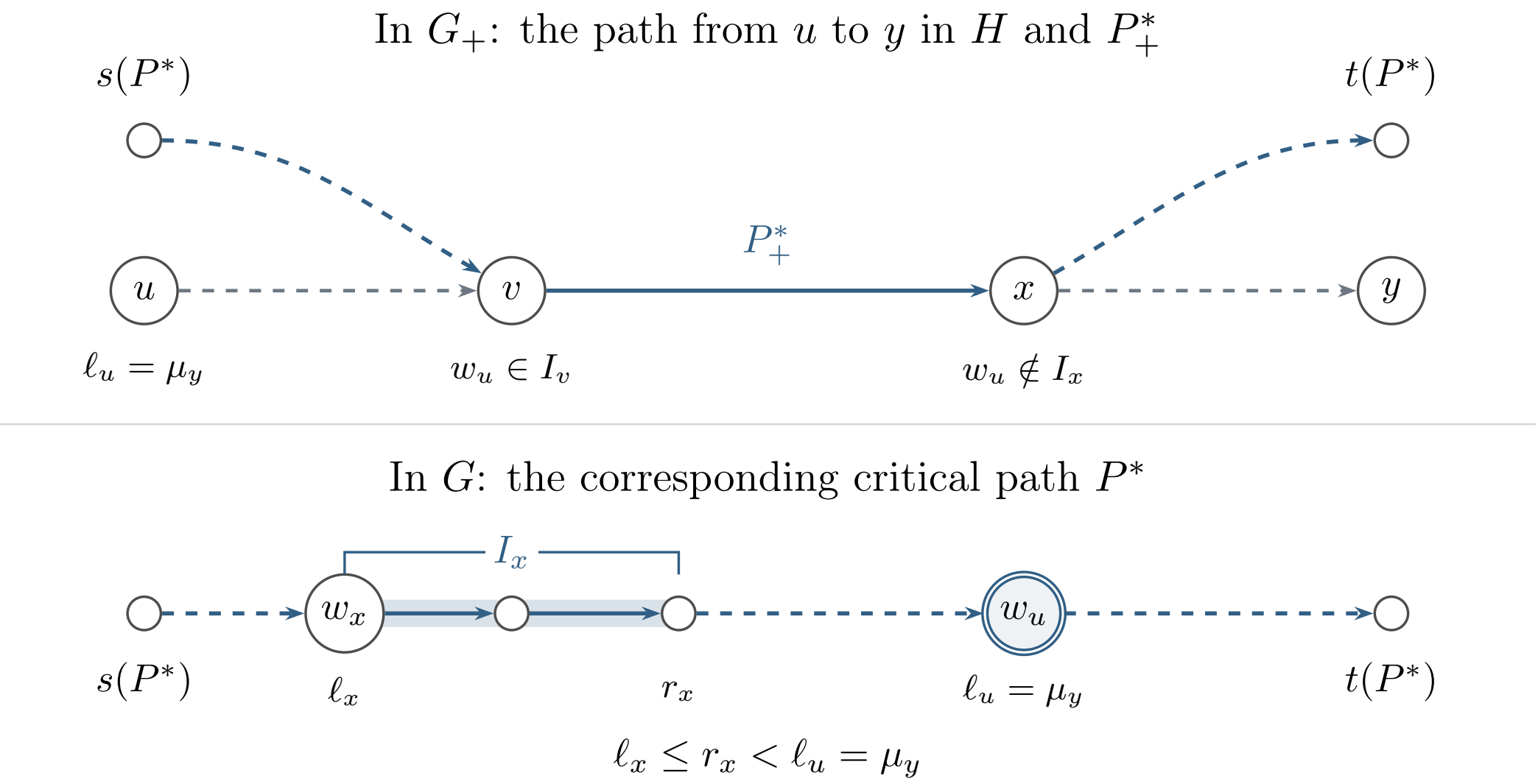

Since all shortcut edges in $G_+$ belong to the support of some $P_+$ with $P \in \calP$, there exists a critical path $P^*$ such that $P^*_+$ contains the edge entering $x$ along the path from $u$ to $y$. Such a $P^*$ then
contains both $w_u$ and $I_x$ since $P^* \in \calF_v, \calF_x$. Additionally, since
$\ell_x\leq\mu_y=\ell_u$ by choice of $u$ and $w_u \not\in I_x$, the interval $I_x$ lies strictly
before $w_u$ on $P^*$. In particular, this implies $I_x \cap I_u = \emptyset$. See \Cref{fig:interval-order-claim}.

However, using $P^*$, we can show that $\calF_u\cup\calF_x$ is a compatible
family. 
\begin{itemize}
    \item Since $x$ is reachable from $u$ in $H$, there is a path $R[u,x]$
    between them in $G_+$. Then, for each $P\in\calF_u$ and
    $Q\in\calF_x$,
    \begin{equation}
    P_+[s(P),u]\circ R[u,x] \circ Q_+[x,t(Q)]  \label{eq:route-1}
    \end{equation}
    is a valid path in $G_+$. Hence, since $G_+$
    preserves reachability, the start of each path in
    $\calF_u$ can reach the end of each path in $\calF_x$
    in $G$.
    \item We also know that $\ell_x\leq\ell_u$ by choice
    of $u$. Let $w_x$ be the first vertex of $I_x$.
    Now, for each $Q\in\calF_x$ and $P\in\calF_u$,
    \begin{equation}
      Q[s(Q),w_x]\circ P^*[w_x,w_u]\circ P[w_u,t(P)]   \label{eq:route-2}
    \end{equation}
    is a valid path in $G$.
\end{itemize}
Now, in every case, we get a contradiction with our initial assumption that $w_u \not\in I_y$. In particular, we use the fact that this assumption implies $I_u \cap I_x = \emptyset$, as we have already observed. 

\begin{itemize}
    \item If $u,x$ are both Steiner vertices,
\Cref{lem:compatible-families-common} immediately gives
$I_u\cap I_x\neq\emptyset$, a contradiction.
\item If exactly one is an original vertex, splice the routes from \Cref{eq:route-1,eq:route-2} at
that vertex: when $u$ is original, this gives a route
$s(Q)\leadsto u\leadsto t(Q)$ for every $Q\in\calF_x$;
when $x$ is original, it gives
$s(P)\leadsto x\leadsto t(P)$ for every $P\in\calF_u$. Here we are using that if $u$ is an original vertex, then $w_u =u$, and similarly for $x$.
Preservation of reachability and uniqueness of critical paths then give
$I_u\cap I_x \neq \emptyset$, a contradiction.
\item Finally, if both are original vertices, $R[u,x]$ and
$P^*[w_x,w_u]=P^*[x,u]$ give mutual reachability in $G$.
Since $G$ is a DAG, this forces $u=x$, also a contradiction.
\end{itemize}
Hence $w_u\in I_y$, giving
$I_u\cap I_y\neq\emptyset$ and $\mu_y\leq r_y$.
\end{proof}
This completes the proof of \Cref{thm:steiner-shortcut-template}.
\end{proof}

\end{document}